\documentclass{article}
\usepackage{graphicx} % Required for inserting images
\newcommand\blfootnote[1]{%
  \begingroup
  \renewcommand\thefootnote{}\footnote{#1}%
  \addtocounter{footnote}{-1}%
  \endgroup
}
\usepackage[a4paper, total={7in, 9in}]{geometry}
\usepackage[
backend=biber,
style=chem-angew,
]{biblatex}
\usepackage{placeins}
\usepackage{wrapfig}

\title{Hierarchical woven fibrillar structures in developing single gyroids in butterflies }

\author{A-L. Jessop$^{a,l,m,*}$, P.L. Clode$^{b,c}$, M. Saunders$^{b}$, M.E. Evans$^{d}$, S.T. Hyde$^{e,f}$, \\ J.N. McPherson$^{g}$, K.S. Pederson$^{g}$, J.J.K. Kirkensgaard$^{h,i}$, N.H. Patel$^{j}$, \\ K.A. DeMarr$^{k}$, W.O. McMillan$^{l}$, B.D.~Wilts$^{m}$, and G.E.~Schr\"oder-Turk$^{a,f,g,h,i}$}

\begin{document}

\maketitle

\section*{Abstract}
Nature offers a remarkable diversity of nanomaterials that have extraordinary functional and structural properties. Intrinsic to nature is the impressive ability to form complex ordered nanomaterials via self-organization. 
One particularly intriguing nanostructure is the Gyroid, a network-like structure exhibiting high symmetry and complex topology. Although its existence in cells and tissues across many biological kingdoms is well documented, how and why it forms remains elusive and uncovering these formation mechanisms will undoubtedly inform bioinspired designs.
A beautiful example is the smooth single gyroid that is found in the wing scales of several butterflies, where it behaves as a photonic crystal generating a vibrant green colour. Here, we report that the gyroid structures of the Emerald-patched Cattleheart, \textit{Parides sesostris}, develop as woven fibrillar structures, disputing the commonly held assumption that they form as smooth constructs. Ultramicroscopy of pupal tissue reveals that the gyroid geometry consists of helical weavings of fibres, akin to hyperbolic line patterns decorating the gyroid. Interestingly, despite their fibrillar nature, electron diffraction reveals the absence of crystalline order within this material. Similar fibrillar structures are also observed in the mature wing scales of \textit{P.\,sesostris} specimens with surgically altered pupal development, leading to a blue colouration. Our findings not only introduce a fundamentally new variation of the gyroid in biology but also have significant implications for our understanding of its formation in nature.

\blfootnote{$^a$School of Mathematics, Statistics, Chemistry and Physics, Murdoch University, 90 South St 6150, Murdoch, Australia.} 
\blfootnote{$^b$Centre for Microscopy, Characterisation, and Analysis, University of Western Australia, 35 Stirling Hwy 6009, Crawley, Australia.} \blfootnote{$^c$School of Biological Sciences, University of Western Australia, 35 Stirling Hwy 6009, Crawley, Australia.} 
\blfootnote{$^d$Institut f\"ur Mathematik, Universit\"at Potsdam, Am Neuen Palais 10, 14469 Potsdam, Germany.} 
\blfootnote{$^e$School of Chemistry, The University of Sydney, Camperdown 2050, Australia.} 
\blfootnote{$^f$Research School of Physics, The Australian National University, Canberra 2600, Australia.}
\blfootnote{$^g$ Department of Chemistry, Technical University of Denmark, Kemitorvet 207, DK-2800 Kgs, Denmark.}
\blfootnote{$^h$ Niels Bohr Institute, University of Copenhagen, Jagtvej 155 A, K{\o}benhavn, 2200, Denmark.}
\blfootnote{$^i$ Department of Food Science, University of Copenhagen, N{\o}rregade 10, K{\o}benhavn, 1172, Denmark.}
\blfootnote{$^j$ Marine Biology Laboratory, University of Chicago,  7 Mbl St 02543, Woods Hole, USA.}
\blfootnote{$^k$ Department of Integrative Biology, University of California Berkeley, 110 Sproul Hall, 5800 Berkeley, USA.}
\blfootnote{$^l$ Smithsonian Tropical Research Institute, Luis Clement Avenue, 0843-03092 Gamboa, Panama.}
\blfootnote{$^m$ Department of Chemistry and Physics of Materials, University of Salzburg, Jakob-Haringer-Str.\,2a, 5020 Salzburg, Austria.}
\blfootnote{*Author for correspondence: A.L. Jessop: annie.jessop@murdoch.edu.au}

\newpage

\section*{Introduction}

\begin{wrapfigure}{r}{0.5\textwidth} 
    \centering
    \includegraphics[width=0.5\textwidth]{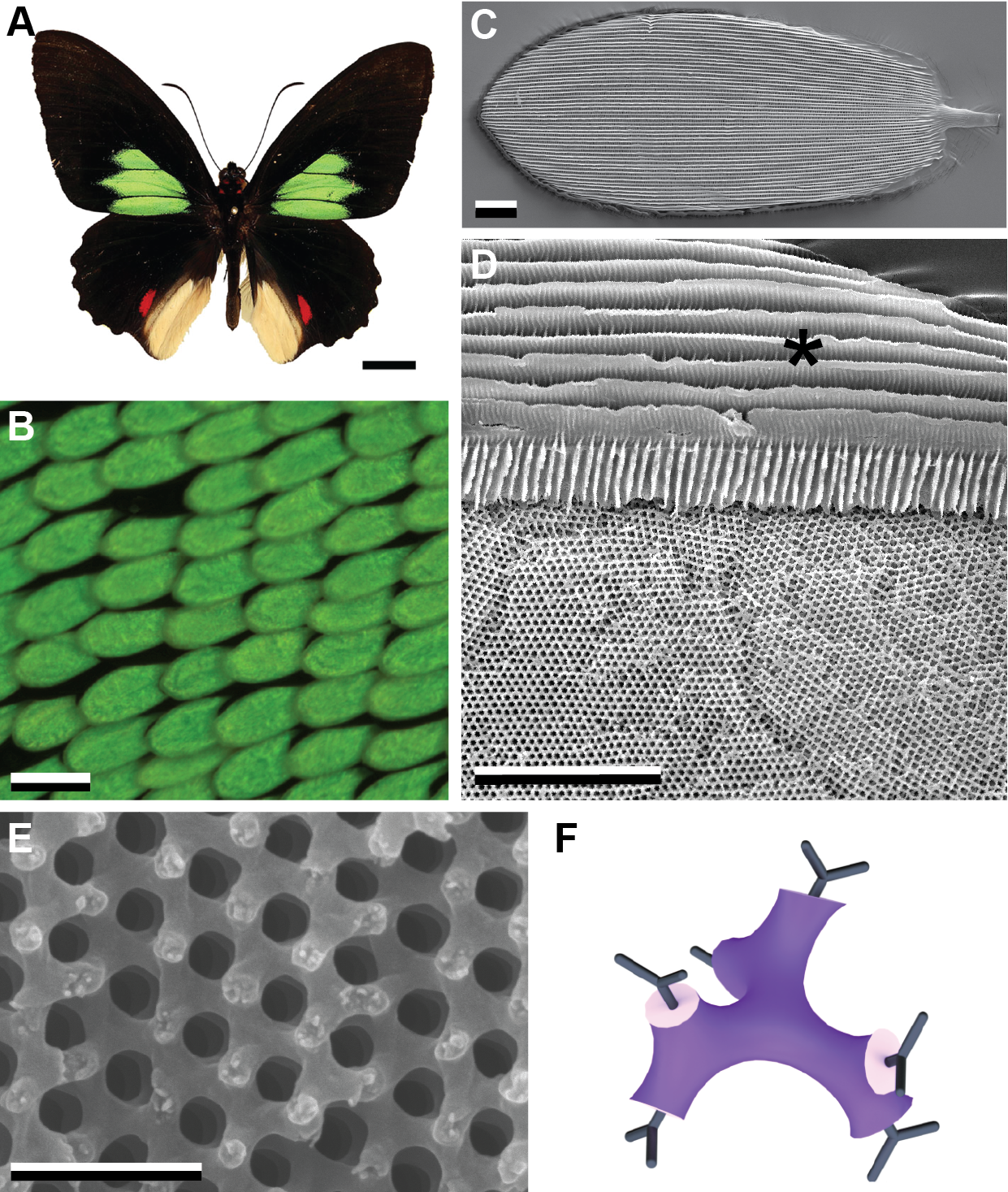}
    \vspace{-16pt}
    \caption{\small{Wing scale design and gyroid structure within the green scales of a mature \textit{P. sesostris} butterfly: (A) Photograph of a male \textit{P. sesostris}. (B) Light microscopy image of gyroid-containing wing scales. (C, D) SEM micrographs of the wing scales which are approximately 100 $\mu$m in length, 50 $\mu$m in width, and 6 $\mu$m thick. The wing scale 'Bauplan' only deviates from the common design of wing scales in that the ridges and cross-rib structure merges into a thick vertical structure that acts as a diffuser, the so-called honeycomb (highlighted with an asterisk in D, \cite{wilts2012iridescence}). (E) High magnification micrograph of a single gyroid nanostructure. With a lattice parameter $a$ of $\approx$ 330--350\,nm, it causes a green reflection and is situated between the diffuser and the continuous lower lamina. (F) Idealised single gyroid nanostructure that is usually modelled to be composed of a solid, homogeneous phase representing chitin --with no internal structure-- and a hollow pore space. The idealised structure uses the nodal approximation to model the chitin domain as the space with $\sin(X)\cos(Y)+\sin(Y)\cos(Z)+\sin(Z)\cos(X)\le -f$ with $f$ adjusted to give a chitin volume fraction of 0.35; $X=2\pi x/a, Y=2\pi y/a, Z=2\pi z/a$. Scale bars: (A) 1 cm, (B) 100 $\mu$m, (C) 10 $\mu$m, (D) 5 $\mu$m, and (E) 500 nm.}}
    \label{fig:standard-view-of-gyroid-in-wingscales}
\end{wrapfigure}

The functional nanostructures that occur in nature often exhibit complex forms that possess extraordinary properties. Of these, the biopolymeric network-like nanostructures that are well-known to occur in butterflies, beetles, and birds are particularly impressive, not least because of their ability to behave as photonic crystals. Particular examples of such structures are the single gyroid nanostructures \cite{michielsen2008gyroid,Saranathan2010,schroder2011chiral,wilts2012iridescence,Saranathan2021}; the single diamond nanostructures  \cite{WiltsDiamondBeetle2012,McNamaraDiamondFossil2014,PouyaOrderedAndDisorderedDiamond2011}, and the disordered forms thereof  \cite{PouyaOrderedAndDisorderedDiamond2011,djeghdi20223d}; the disordered (scattering) structures of white beetles \cite{wilts2018evolutionary}; ordered and disordered versions of the I-WP structure \cite{bauernfeind2024not,kobayashi2021discovery}; and structures reminiscent of arrested spinodal decomposition in birds \cite{dufresne2009self}. While there is substantial diversity in how these nanostructures are embedded within the organisms, the porous structures share a common design: a single component of solid material has a network-like geometry, embedded in air. The pore space, the complement of the solid material, is also a single connected component, cf.\ Fig.\ \ref{fig:standard-view-of-gyroid-in-wingscales}. In the case of the ordered single gyroid and single diamond structures, the pore and solid space are of identical topology and symmetry, albeit occupying different volume fractions.

One question that has fascinated biologists and physicists alike is the formation mechanism of these network-like nanostructures \cite{Saranathan2010, michielsen2008gyroid, wilts2019nature, ghiradella1989structure}. For different network-like structures in different organisms, formation mechanisms from self-assembly, to templating \cite{Saranathan2010, WiltsScienceAdv2017, ghiradella1989structure}, to bottom-up phase separation \cite{Saranathan2021}, or combinations of these have been proposed. However, little progress has been made to determine the formation mechanisms experimentally in developing organisms.

Notably, in all electron microscopic analyses of these photonic ordered network-like structures to date, the solid biopolymeric material constituting the gyroid or diamond structures has no discernible textural features. This is in contrast to the distinct textural features observed in micrographs of the Bouligand ‘helical plywood’ structure of beetles, which are closely related to the cholesteric phase in liquid crystals and similar to other fibrillar assemblies, such as in cellulose \cite{ParkerVignoliniCelluloseSelfAssAdvancedMater2018,wilts2014natural}. While the role of microfilaments and other fibres in butterfly wing scale development has long been recognised (see \cite{Ghiradella1974,Dinwiddie2014}), no fibrillar structure has been discussed in relation to single gyroid nanostructures. 

Additionally, these network-like structures are often assumed to be predominately chitin. Biopolymers, including chitin, are known to occur in different crystalline and amorphous forms and we have detailed knowledge of naturally occurring chitin phases in crustaceans, mushrooms, and beetles \cite{hou2021understanding}. The phases of related materials such as chitosan or synthetically reconstituted chitin nanofibres are also well described. While some studies have investigated the crystallinity of butterfly wings \cite{schiffman2009solid, binetti2009natural}, no studies have specifically addressed the chitin phase in the nanostructures responsible for structural colour in butterflies including the network-like structures investigated here.

Here, we describe the structural forms of the single gyroid photonic nanostructures of \textit{P. sesostris} butterflies from different stages of development using electron microscopy. Remarkably, at all investigated stages beyond 60\% development, the forming scales within the pupae exhibit a novel biological structure: a single gyroid nanostructure composed of woven fibres. Furthermore, when the pupa was partially dissected, mechanically damaged, and exposed to antibiotics at an early stage of development, the woven structure persisted in affected wing scales even within the mature and sclerotised scales. We rationalise the observed structure using a geometric model that constructs the gyroid from disconnected woven helices. We also use Electron Diffraction (ED) and Wide- and Small-angle X-ray Diffraction (WAXS / SAXS) to determine the crystallographic properties of fully formed adult single gyroid nanostructures and the woven fibrillar structures observed in developing pupae and discuss the implications of these findings for the growth models of gyroid-forming butterflies.

\section*{Results}
Our key finding is that, in the early stages of its formation, the single gyroid structure that forms in the wing scale cells of \textit{P.\,sesostris} pupae are composed of a complex array of entangled fibres. From our electron micrographs, it appears that the interlocking twists of these fibrils wind symmetrically around the \texttt{srs} graph (the skeletal graph representing a single gyroid, also known as the Laves graph \cite{SabaPRL2011}).

\begin{figure}[h]
    \centering
    \includegraphics[width=\textwidth]{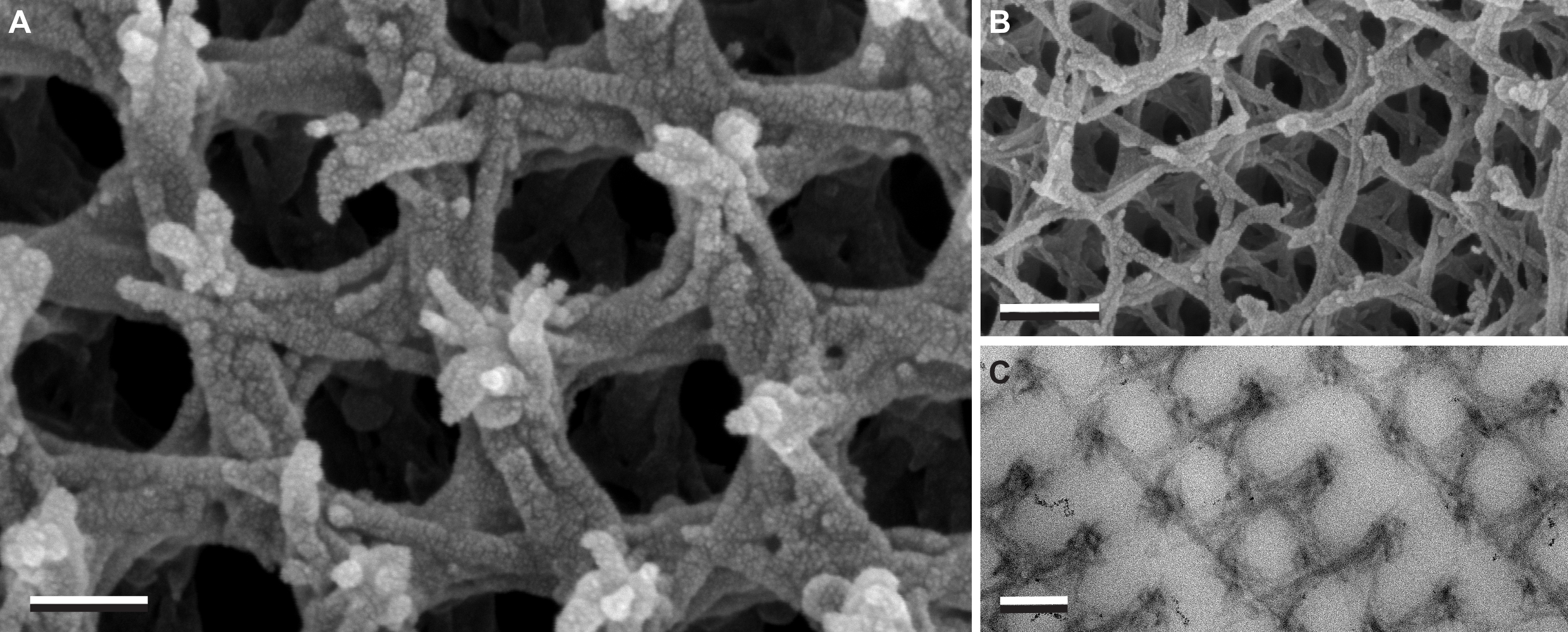}
    \vspace{-16pt}
    \caption{\small{Electron microscopy images of the fibrillar gyroid structures found in the wing scales of developing \textit{P.\,sesostris} pupae. Wing scale development was arrested via fixation at day 13 (70\% development; A) and day 15 (80\% development; B,C), of an approximate 18 day pupation period. (A) A fibrillar gyroid structure adjacent to the lower lamina of the wing scale from a critical-point dried sample shows approximately six entangled fibres making up the gyroid structure. (B) The fibrillar gyroid structure closer to the upper lamina of the wing scale from a critical-point dried sample shows two--three entangled fibres. (C) A TEM section of a wing scale sampled from the same wing displayed in (B) shows that the fibrillar nature of the gyroid structure can also be observed in cross-section. The structures shown here are from individuals that, up to the fixation on day 13 and 15, respectively, have developed in natural conditions without any chemical or mechanical intervention or dissection. Scale bars: (A) 100 nm, (B) 200 nm, and (C) 200 nm.}}
    \label{fig:parides-development-fibres}\label{fig:parides-development-fibers}
\end{figure}

Figure \ref{fig:parides-development-fibres} illustrates this finding at two different developmental stages: at approximately\ 70\% (Figure \ref{fig:parides-development-fibers}A) and\ 80\% development (Figure \ref{fig:parides-development-fibres}B, C). The fibrillar organisation is evident in both SEM images of critical-point dried samples (Figure \ref{fig:parides-development-fibres}A, B) and in TEM images (Figure \ref{fig:parides-development-fibres}C) of a scale that was sampled from the same wing as the sample presented in Figure \ref{fig:parides-development-fibres}B. 

\newpage

The fibres are approximately 25\,nm in diameter and, depending on the developmental stage and location within the wing scale lumen, two to fifteen fibres are woven (based on estimates from strut diameter, assuming a densely packed array of fibres). Earlier in development, fewer fibres appear to be woven, compared to later stages (Figure S1A--C). Additionally, we observe that the structure closest to the lower lamina (Figures\ \ref{fig:parides-development-fibres}A,\,S1E) contains more entangled fibres compared to the structure closest to the upper lamina (Figure\ \ref{fig:parides-development-fibres}B,\,S1D). We also observe a fibrillar morphology in the upper lamina's `honeycomb' diffuser structure that is located between the scale ridges and the gyroid structure in the developing scales (see Figures S1D and S2D).

Importantly, this fibrillar structure of the gyroid during the pupal development contrasts strongly to the homogeneous smooth nanostructure that is observed in adult wing scales of the same \textit{P.\,sesostris} population (Figure\ \ref{fig:standard-view-of-gyroid-in-wingscales}E).

\begin{figure}[h!]
    \centering
    \includegraphics[width=\textwidth]{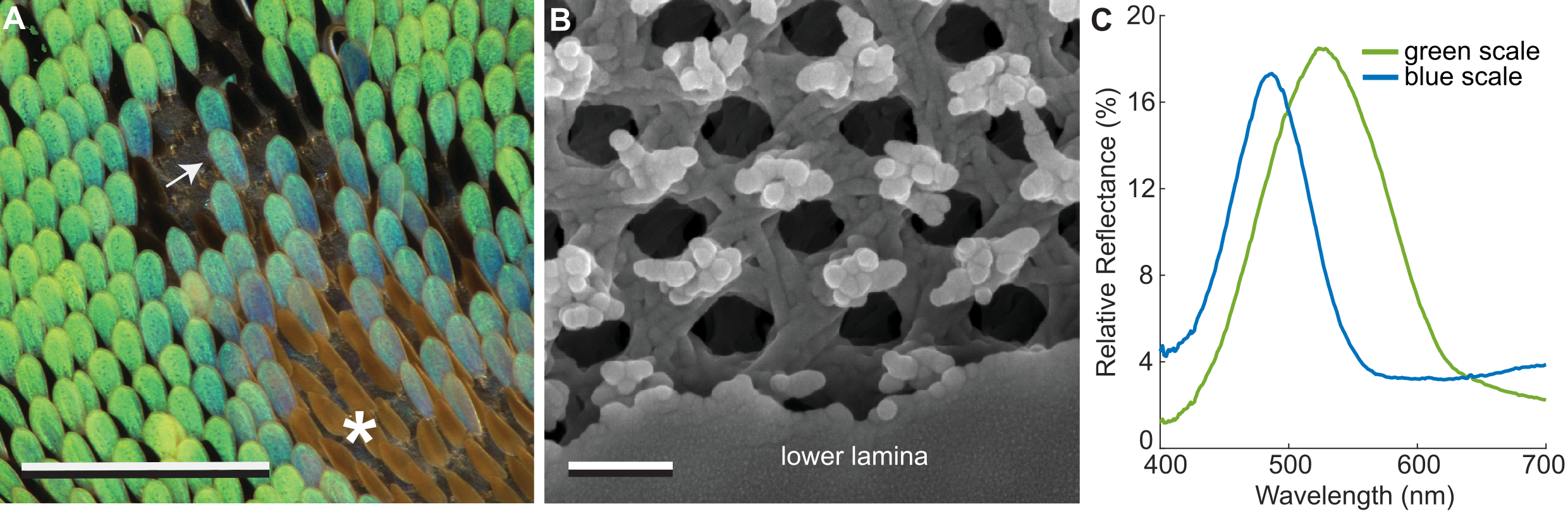}
    \vspace{-16pt}
    \caption{\small{Wing scales from a \textit{P. sesostris} butterfly that underwent surgery for \textit{in vivo} imaging involving dissecting part of the pupal cuticle to expose the developing wing. During the dissection, part of the wing suffered an injury resulting in deformed wing scales that did not fully develop. (A) Light microcopy image of the injury site showing normal green cover scales and black ground scales, blue wing scales (indicated by the arrow) and orange scales (indicated by the asterisk). The blue wing scales are gyroid-containing scales and the orange scales are the black ground scales that abnormally developed due to the dissection. (B) SEM image showing the fibrillar structure of the gyroid photonic crystal in the blue wing scales. (C) Relative reflectance measurements of the blue and green wing scales depicted in (A) measured with microspectrophotometry. Peak reflectances for the green wing scales occurs at approximately 535 nm and for the blue wing scales at approximately 490 nm. Scale bar = 500 $\mu$m (A) and 200 nm (B).}}
    \label{fig:parides-mature-but-dissected-blue-scales}
\end{figure}

Further, the fibrillar nature of the gyroid nanostructure persists in the mature wing scales in specimens whose development was stunted by a surgical procedure (see Methods). In these butterflies, gyroid-forming wing scales near the incision line develop abnormally and exhibit a blue structural colour rather than the usual green in the mature butterfly. Using microspectrophotometry, we show that this abnormal development results in an approximate 40 nm blue shift in peak reflectance between the green and blue scales (Figure \ref{fig:parides-mature-but-dissected-blue-scales}C). Additionally, the ground scales appear brown rather than black (Figure \ref{fig:parides-mature-but-dissected-blue-scales}A). As shown in Figure \ref{fig:parides-mature-but-dissected-blue-scales}B, the gyroid nanostructure in these mature blue scales is retained and has a similar fibrillar appearance to the fibrillar nanostructure observed at the early developmental stages in normally developed specimens (Figure \ref{fig:parides-development-fibres}).

\vspace{-6pt}

\begin{figure}[h]
    \centering
    \includegraphics[width=14cm]{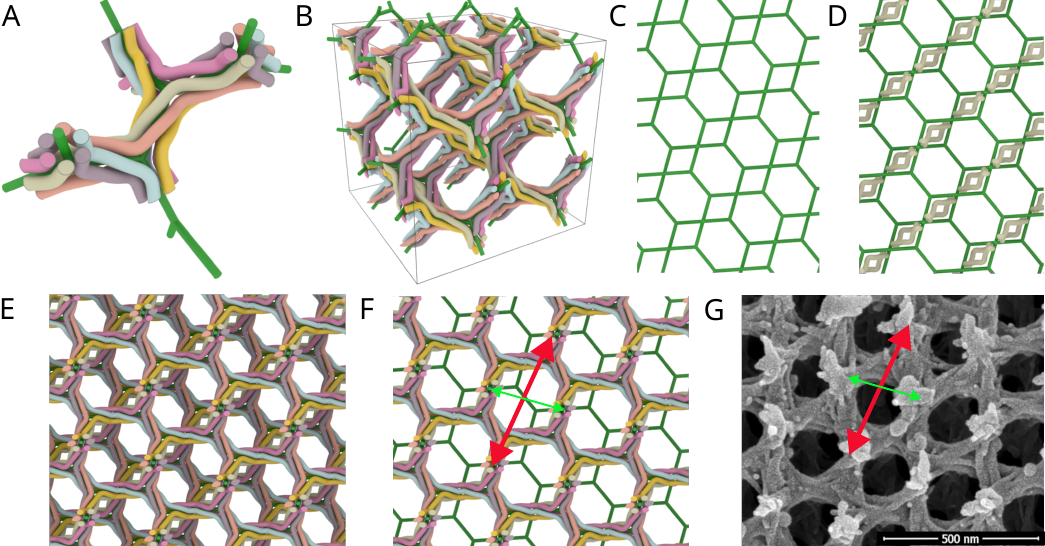}
    
    \caption{\small{Geometric model for a 6-fold helical filamentous weave around the {\tt{srs}} (Gyroid) network. This model corresponds to an entangled assembly of individual (deformed) helices with helix axes aligned with the six different [110] directions. The \texttt{srs} graph is shown in green; the filaments are coloured by the direction of the symmetry screw axis around which they revolve (of which there are six). (A) A small section of the model; (B) Approximately $2^3$ unit cells; (C) [110] projection of the \texttt{srs} graph; (D) As in C, with one set of parallel filaments; (E) A half-space of the full six-fold weaving, clipped by a [110] plane, similar to the termination plane in the butterfly; (F) Same as E, except the model is confined to a thin sheet rather than a half-space to mimic the finite observation depth of SEM micrograph (the \texttt{srs} graph is shown in full); in this thin slice, the gyroidal woven structure has a hexagonal appearance that is also observed in the SEM image of the butterfly woven structure with the clipping plane corresponding to the [110] direction. (G) An SEM image of a developing \textit{P.\,sesostris} gyroid at day 13 (70\% development) demonstrating the visual similarity to the model in F.}} 
    \label{fig:weaving-model}
\end{figure}

There is a geometric connection between symmetric minimal surfaces, fibrils, and entangled geometries. Entangled fibrillar structures can be considered as combinatorial objects on a scaffold network, in this case the \texttt{srs} network, which is the labyrinth graph of the single gyroid. These structures are analogous to highly symmetric entanglements wound on polyhedral and honeycomb structures \cite{hyde2022symmetric,evans2022symmetric}. We discuss two particular models with six fibrils: six fibrils with a twist of $\frac{1.8}{6}*2\pi$, which we call $(\frac{1.8}{6})^6$, and six fibrils with a twist of $\frac{0.8}{6}*2\pi$, which we call $(\frac{0.8}{6})^6$. These 6-fold helical models resemble the fibrillar structure observed in some of the butterfly scales above (see Figure \ref{fig:weaving-model}). From a global perspective, the structures resemble two ideal, related weavings composed of filaments wound around edges of an \texttt{srs} network: (i) arrangements of non-overlapping, entangled helices wound around the [110] axes ($(\frac{0.8}{6})^6$, which is equivalent to structure $G^{+}_{118RL}(cosh^{-1}(\sqrt{6}))$ (Fig. 16 in  \cite{evans2013periodic}) as shown in Figure \ref{fig:weaving-model}) and (ii) triplets of helices ($(\frac{1.8}{6})^6$, see Figure S3) wound around the cubic [111] axes, forming the $\Sigma^+$ rod packing \cite{okeeffe2001invariant}. Slices normal to the [110] direction from either model resemble the fibrillar structure observed in the butterfly scales (c.f.\, Figure \ref{fig:weaving-model}E \& F). 

Further, we used ED and SAXS/WAXS to assess the crystallinity of the materials that comprise the gyroid nanostructures. ED of TEM sections of developing scales gave no indication of crystalline order in the fibrous nanostructures at 80\% development (Figure S2 and S4). ED of green scales from a mature butterfly wing that visually appeared to contain exclusively gyroid structures was characteristic of atomic scale disorder (see Figure S7). Only in a very small number of scales did we observe fleeting diffraction that may or may not have been caused by the gyroid; these could not be clearly indexed to a chitin phase (see Figure S8).

SAXS/WAXS analyses of concentrated and subsequently dessicated suspensions of mature \textit{P.\ sesostris} wing scales, found no indication of structural order associated with chitin (Figure S5 and S6). This is in contrast to similarly prepared samples of \textit{Morpho portis} scales and in contrast to whole wing samples of \textit{P.\ sesostris}; both show faint but visible scattering related to chitin relatives (see Figure S5 and S6).

\vspace{-10pt}

\section*{Discussion}
\subsection*{Fibrillar nature of developing Gyroid nanostructure}

It is commonly assumed that gyroid nanostructures in butterflies form homogeneous solid structures with smooth interfaces, based on observations of fully developed wing scales \cite{WiltsScienceAdv2017, michielsen2008gyroid, Saranathan2010, schroder2011chiral}. However, our findings challenge this assumption. Rather, the initial nanostructure is an intricate entanglement of nanofibres, each $\approx 25$ \,nm in diameter (Figure \ref{fig:parides-development-fibers}). 

Chitin, the key component of insect cuticle, is known to polymerise into microfibrils \cite{merzendorfer2003chitin}, and such fibrillar structures are widespread across biopolymers, however typically with less than half the diameter we observe in \textit{P. sesostris}. For instance, chitin microfibrils have been described in fungi \cite{bartnicki1978isolation, ifuku2011preparation} and crustaceans \cite{hou2021understanding}, while keratin and cellulose-containing structures have been found in birds \cite{wang2016keratin} and plants \cite{saxena2005cellulose}, respectively. Additionally, ultra-microscopy studies of butterflies have revealed fibrillar elements in developing wing scale cells, mostly as actin filaments \cite{lloyd2024actin,seah2023hierarchical,day2019sub,Dinwiddie2014,McDougalKolle2021PNAS} but also as nascent cuticle fibrils, e.g.\ in developing \textit{Celastrina ladon} \cite{ghiradella1989structure}. Although fibrillar structures are abundant in nature, those that appear to construct a gyroid from multiple entangled fibres, like the structure that we describe here, are so far, extremely rare. A similar but different filamentous gyroidal woven structure has also been suggested to form in the strateum corneum, the outer layer of mammalian skin, albeit at a much smaller length scale (lattice parameter $\le 30$ nm \cite{NORLEN2004}).  

Theoretical models further support the plausibility of gyroid formation via fibres. For instance, gyroid-like structures can emerge from threaded or overlapping helices \cite{wetzel2024triplyperiodichelicalweaves,Hielscher:17} or dense hyperbolic packings of congruent filaments projected onto triply-periodic minimal surfaces (TPMS) \cite{evans2013periodic}. In addition, simulations of self-assembled mixtures of star copolymers have produced network structures on TPMS that resemble these biological forms  \cite{KirkensgaardPNASStripedGyroids2014}. The gyroidal weaving in skin cells \cite{NORLEN2004}) has been idealised as a rod packing with the same symmetry group as the single gyroid and its rod-to-helix deformation modes may be responsible for the ability of skin to swell \cite{EvansRothPRL2014Skin,EvansHydeSkin2011}. 

We have shown that the fibrous gyroid of \textit{P. sesostris} is structurally amorphous at the atomic scale. Electron diffraction gave no evidence of crystallinity in the developing nanostructure (Figures S2 and S4). This is notable because, as stated above, chitin fibres are known to occur widely across different organisms, typically in one of three different crystalline modifications: $\alpha$-, $\beta$- and/or $\gamma$-chitin \cite{hou2021understanding}. Non-crystalline transient states of chitin have been shown to occur in certain species of fungi \cite{vermeulen1986chitin} but have rarely been reported among naturally occurring polymers.  Our ED analysis motivates further research addressing the crystallographic properties and elemental composition of butterfly nanostructures, and the interplay of chitin and other cuticle components in stabilising complex nanostructures.

\subsection*{Implications for formation mechanisms}

The formation of biophotonic gyroid nanostructures in butterflies has been a topic of sustained interest, and much of the current understanding stems from the seminal work of Helen Ghiradella on developing \textit{Callophrys gryneus} (formerly \textit{Mitoura gryneus}) wing scales  \cite{ghiradella1989structure}. In this species, the gyroid nanostructures appear as discrete, discontinuous crystallites, similar to that of \textit{Erora opisena} \cite{WiltsScienceAdv2017}. Ghiradella observed that the lumen of developing wing scales contained arrays of 'membrane-cuticle' units, comprising a cylindrical sleeve of membrane surrounding a core of nascent cuticle. These membranes appear to be invaginations of the plasma membrane, so that the enclosed spaces of the units are extracellular and therefore chitin deposition can occur extracellularly, as is typical of cuticle deposition in insects.

Building on these observations, Saranathan \textit{et al.} \cite{Saranathan2010} described a pentacontinuous double gyroid structure underlying the structure formation, encompassing the plasma membrane, smooth endoplasmic reticulum (SER) membrane, extracellular space, intracellular space, and intra-SER space. The model suggests that this pentacontinuous structure forms via self-assembly creating a template for nascent chitin that is deposited into the extracellular space \cite{Saranathan2010}. Once scale development is complete, the wing scale cell dies and recedes leaving behind a solid single gyroid network. 

Our findings suggest an additional stage of development during wing scale growth: rather than chitin deposition occurring as a homogenous extrusion process and closely following the membranous template, single fibres (of currently unknown length) are deposited by the scale cell and are woven into the observed gyroid. This fibrillar formation does not preclude membrane-templating in some form but reveals an extra stage in the process. This observation raises a number of new questions on structure formation in butterflies. In particular, the process whereby the fibrous structure is eventually smoothed into the homogenous solid structure observed in fully developed adult butterflies remains unknown. Possibly, the smoothing is effected by the addition of pigments and sclerotization agents that are embedded within the structure very late in development, shortly before eclosion \cite{iwata2014real}. It is, therefore, possible that the fibrous samples in the mature abnormal scales did not undergo all scale-forming steps, including the addition of pigments \cite{matsuoka2018melanin}. This would also explain the different optical and structural appearances of the disturbed scales of Figure~\ref{fig:parides-mature-but-dissected-blue-scales}, where the blue colouration likely arises from a different filling fraction, length scale, and/or different pigmentation. Pigments are known to be embedded within the wing scales of gyroid-forming butterflies \cite{WiltsScienceAdv2017,wilts_pigmentary_2014,saba2014absence}, but exactly what influence they have on the final morphology of the nanostructure is as yet unknown. Alternatively, the smoothing process may involve chitin degradation by chitinases, which are crucial for insect growth and morphogenesis by remodeling chitin-containing structures \cite{merzendorfer2003chitin}. However, whether chitinases in wing scale cell plasma membranes actively remodel these structures remains unclear. As a further alternative, the smoothing process may result from the rapid condensation of filamental or tubular elements, as observed in developing thin-film laminae of \textit{Celastrina ladon} \cite{ghiradella1989structure}.

\subsection*{Conclusion}

Our discovery of a fibrillar intermediate stage during scale formation changes the paradigm for how complex network-like nanostructures form in butterflies and insects. Firstly, it disproves the widespread assumption that gyroid formation in the butterfly is a smooth extrusion process of amorphous chitin into a membrane-template and secondly, it raises key questions about how these nanofibres are transformed into the smooth, solid gyroids of mature scales. Demonstrating that a weaving of fibres into the gyroid is not only a theoretical possibility, but a reality in an experimental biological system, will undoubtedly inspire new research across biological, synthetic, and bioinspired material systems.

\section*{Methods}

\subsection*{\textit{P. sesostris} pupae and wing dissections}
Wild male and female \textit{P. sesostris} were caught with a butterfly net in Gamboa, Panama and transported to the butterfly rearing facilities at the Smithsonian Tropical Research Institute in Gamboa, Panama. Butterflies reproduced within the facility and eggs were collected daily and transferred to smaller rearing cages. Once hatched, larvae were reared on \textit{Aristolochia sp.} and monitored until pupation. The day before pupation \textit{P. sesostris} larvae secure themselves upright to their host plant. The date that this occurred for each larva was recorded as day zero, and the day that pupation occurred was recorded as day one of pupal development. The percentage of development for individuals that were euthanized after whole wing dissections (N = 4) was estimated using the average duration of pupal development of other \textit{P. sesostris} pupae (18.4 days, N = 5). Pupae were immobilized using a malleable modelling compound (Play-Doh). A micro knife (10315-12, Fine Science Tools) was then used to make an incision into the cuticle around the region of the wing. The cuticle was then removed and the whole wing was exposed. The forewing was cut at the wing base and placed into a 2\% paraformaledhyde/2\% glutaraldehyde in phosphate buffered solution (PBS) for fixation. 

\subsection*{SEM imaging of pupal and adult wing scales}
Fixed pupal wing tissue was dissected into small 1 $mm^2$ pieces from the region of the wing that develops scales with photonic gyroid nanostructures. This tissue was then dehydrated through a graded series of ethanols in a PELCO Biowave microwave processor (see supplementary methods) before being critical-point dried. Scales from adult wings and dried pupal wing samples were gently removed using a cotton swab and placed onto Cu tape on an aluminium stub before being coated with 4 nm Pt. Imaging was conducted on a Thermo Fisher Scientific Verios XHR SEM at 5 kV using the in-lens TLD detector. 

\subsection*{TEM imaging and Selected Area Diffraction of developing wing scales}
Fixed pupal wing tissue was dissected into small 1\,$mm^2$ pieces from the region of the wing that develops scales with photonic gyroid nanostructures. These tissue samples were then post-fixed in 1\% OsO$^4$ in PBS and dehydrated through a graded ethanol series, before being transferred to anhydrous acetone. Infiltration was conducted through a graded series of Procure-araldite resin:acetone mixtures, before being embedded in 100\% resin and cured for 48 h at 70$^o$C. All steps prior to embedding were undertaken in a PELCO Biowave microwave processor (see supplementary methods). For routine TEM imaging and selected area electron diffraction (SAED), sections $\approx$250 nm-thick were prepared using an Ultra 45$^o$ diamond knife on a Leica UC6 ultramicrotome and collected on naked Cu grids. Both imaging and SAED was conducted at 200 kV on a JEOL JEM-F200-HR FEGTEM fitted with a Gatan OneView camera. For SAED, small areas of sectioned scales (e.g. lower lamina, gyroid) were isolated using an appropriately sized selected area aperture.

\subsection*{Electron diffraction of \textit{mature} wing scales from normally developed butterflies}
Green scales from mature butterflies were loaded, either whole or after being gently crushed between two glass slides, onto a continuous carbon film transmission electron microscopy grid, which was then transferred into a Rigaku XtaLAB Synergy-ED dedicated electron diffractometer, equipped with a Rigaku Oxford Diffraction HyPix-ED detector \cite{Ito2021}. Data acquisition was performed at ambient temperature with an electron wavelength of 0.0251\AA\ (200 kV). The eucentric height of the stage was adjusted so that the grains were centred inside the selected area aperture, and then rotated (typically through $-60^o \le \alpha \le 60^o$) continuously. The data were processed using CrysAlisPro \cite{Rigaku}.

\subsection*{SAXS/WAXS}
X-ray scattering analysis was carried out on powder samples of 10,000\,s of adult wing scales of \textit{P.\ sesostris} and \textit{Morpho portis} (scraped off dry adult specimens using cotton buds and ethanol, and concentrated in a centrifuge in NMR tubes),  intact wing segments with scales of \textit{P.\ sesostris}, as well as elytral fragments of the beetle \textit{Anomala cupripes} as reference. Measurements used a Nano-inXider instrument (Xenocs SAS, Grenoble, France) with a 40 W micro-focused Cu source producing X-rays with a wavelength of $\lambda=1.54$ \AA. The measurements were performed at a medium resolution configuration (50 kV, 0.6 mA, 800 $\mu$m beam size, $\approx$ 80 MPh/s flux), with 1--8 hours measurement time in vacuum.

\subsection*{Microspectrophotometry}
A modified Zeiss Axioscope 5 optical light microscope was used to perform microspectrophotometry (MSP) measurements. A Zeiss EC Epiplan-NEOFLUAR objective with a magnification of 50$\times$ (NA = 0.55) was used for the measurement. Illumination was provided by a halogen light source (HAL 100, Zeiss). Relative reflectance spectra were collected via microscope sideport below the tube lens  with the light path consisting of a mirror, a focusing quartz lens, and a 600 \textmu m quartz fibre (FC-UVIR600-2-BX, Avantes) attached to a spectrometer (AvaSpec-ULS2048XL-EVO, Avantes). An aluminium mirror (PF10-03-F01, Thorlabs) served as the reference.

\section*{Acknowledgements}
This work was supported by the Australian Research Council (ARC) through the Discovery Project DP200102593 and a research grant from the Human Frontiers in Science Program (ref.-no:
RGP0034/2021 to B.D.W.). X-ray scattering data were generated using a research infrastructure at the University of Copenhagen, partly funded by FOODHAY (Food and Health Open Innovation Laboratory, Danish Roadmap for Research Infrastructure). The authors acknowledge the facilities and staff of Microscopy Australia at the Centre for Microscopy, Characterisation and Analysis at The University of Western Australia. In particular, we thank Drs Crystal Cooper, Alexandra Suvorova, and Samantha Gunasekera for their fantastic support in both sample preparation and imaging. We also thank R\'{e}mi Mauxion, Abby Williams, Leo Camino, Oscar Paneso, and staff at the Smithsonian Tropical Research Institute for their support in rearing the butterflies.

\printbibliography

@article{Tsurkan2021,
  title = {Progress in chitin analytics},
  volume = {252},
  ISSN = {0144-8617},
  url = {http://dx.doi.org/10.1016/j.carbpol.2020.117204},
  DOI = {10.1016/j.carbpol.2020.117204},
  journal = {Carbohydrate Polymers},
  publisher = {Elsevier BV},
  author = {Tsurkan,  Mikhail V. and Voronkina,  Alona and Khrunyk,  Yuliya and Wysokowski,  Marcin and Petrenko,  Iaroslav and Ehrlich,  Hermann},
  year = {2021},
  month = jan,
  pages = {117204}
}

@article{saxena2005cellulose,
  title={Cellulose biosynthesis: current views and evolving concepts},
  author={Saxena, Inder M and Brown Jr, R Malcolm},
  journal={Annals of Botany},
  volume={96},
  number={1},
  pages={9--21},
  year={2005},
  publisher={Oxford University Press}
}

@Article{evans2022symmetric,
AUTHOR = {Evans, Myfanwy E. and Hyde, Stephen T.},
TITLE = {Symmetric Tangling of Honeycomb Networks},
JOURNAL = {Symmetry},
VOLUME = {14},
YEAR = {2022},
NUMBER = {9},
ARTICLE-NUMBER = {1805},
URL = {https://www.mdpi.com/2073-8994/14/9/1805},
ISSN = {2073-8994}
}

@article{hyde2022symmetric,
author = {Stephen T. Hyde  and Myfanwy E. Evans },
title = {Symmetric tangled Platonic polyhedra},
journal = {Proceedings of the National Academy of Sciences},
volume = {119},
number = {1},
pages = {e2110345118},
year = {2022},
doi = {10.1073/pnas.2110345118},
URL = {https://www.pnas.org/doi/abs/10.1073/pnas.2110345118}
}

@manual{Rigaku,
    key = "Rigaku Oxford Diffraction",
    title = "CrysAlisPro software system version 1.171.44.88a, Rigaku Corporation, Wroclaw, Poland", 
    year = 2024
}

@article{evans2013periodic,
author = "Evans, Myfanwy and Robins, Vanessa and Hyde, Stephen",
title = "{Periodic entanglement II: weavings from hyperbolic line patterns}",
journal = "Acta Crystallographica Section A",
year = "2013",
volume = "69",
number = "3",
pages = "262--275",
doi = {10.1107/S0108767313001682},
url = {https://doi.org/10.1107/S0108767313001682}
}

@article{okeeffe2001invariant,
author = "O'Keeffe, M. and Pl{\'{e}}vert, J. and Teshima, Y. and Watanabe, Y. and Ogama, T.",
title = "{The invariant cubic rod (cylinder) packings: symmetries and coordinates}",
journal = "Acta Crystallographica Section A",
year = "2001",
volume = "57",
number = "1",
pages = "110--111",
doi = {10.1107/S010876730001151X},
url = {https://doi.org/10.1107/S010876730001151X}
}

@article{wang2016keratin,
  title={Keratin: Structure, mechanical properties, occurrence in biological organisms, and efforts at bioinspiration},
  author={Wang, Bin and Yang, Wen and McKittrick, Joanna and Meyers, Marc Andr{\'e}},
  journal={Progress in materials science},
  volume={76},
  pages={229--318},
  year={2016},
  publisher={Elsevier}
}

@Article{Ito2021,
author ="Ito, Sho and White, Fraser J. and Okunishi, Eiji and Aoyama, Yoshitaka and Yamano, Akihito and Sato, Hiroyasu and Ferrara, Joseph D. and Jasnowski, Michał and Meyer, Mathias",
title  ="Structure determination of small molecule compounds by an electron diffractometer for \uppercase{3D ED/M}icro\uppercase{ED}",
journal  ="CrystEngComm",
year  ="2021",
volume  ="23",
issue  ="48",
pages  ="8622-8630",
publisher  ="The Royal Society of Chemistry",
doi  ="10.1039/D1CE01172C",
url  ="http://dx.doi.org/10.1039/D1CE01172C"}

@article{kobayashi2021discovery,
  title={Discovery of \uppercase{I-WP} minimal-surface-based photonic crystal in the scale of a longhorn beetle},
  author={Kobayashi, Yuka and Ohnuki, Ryosuke and Yoshioka, Shinya},
  journal={Journal of the Royal Society Interface},
  volume={18},
  number={184},
  pages={20210505},
  year={2021},
  publisher={The Royal Society}
}

@article{matsuoka2018melanin,
  title={Melanin pathway genes regulate color and morphology of butterfly wing scales},
  author={Matsuoka, Yuji and Monteiro, Ant{\'o}nia},
  journal={Cell reports},
  volume={24},
  number={1},
  pages={56--65},
  year={2018},
  publisher={Elsevier}
}

@article{wilts2014natural,
  title={Natural helicoidal structures: morphology, self-assembly and optical properties},
  author={Wilts, Bodo D and Whitney, Heather M and Glover, Beverley J and Steiner, Ullrich and Vignolini, Silvia},
  journal={Materials Today: Proceedings},
  volume={1},
  pages={177--185},
  year={2014},
  publisher={Elsevier}
}

@article{djeghdi20223d,
  title={3D Tomographic Analysis of the Order-Disorder Interplay in the \textit{Pachyrhynchus congestus mirabilis} Weevil},
  author={Djeghdi, Kenza and Steiner, Ullrich and Wilts, Bodo D},
  journal={Advanced Science},
  volume={9},
  number={26},
  pages={2202145},
  year={2022},
  publisher={Wiley Online Library}
}

@article{ifuku2011preparation,
  title={Preparation of chitin nanofibers from mushrooms},
  author={Ifuku, Shinsuke and Nomura, Ryoki and Morimoto, Minoru and Saimoto, Hiroyuki},
  journal={Materials},
  volume={4},
  number={8},
  pages={1417--1425},
  year={2011},
  publisher={MDPI}
}

@article{bartnicki1978isolation,
  title={Isolation of chitosomes from taxonomically diverse fungi and synthesis of chitin microfibrils in vitro},
  author={Bartnicki-Garcia, Salomon and Bracker, Charles E and Reyes, Emma and Ruiz-Herrera, Jos{\'e}},
  journal={Experimental Mycology},
  volume={2},
  number={2},
  pages={173--192},
  year={1978},
  publisher={Elsevier}
}

@article{merzendorfer2003chitin,
  title={Chitin metabolism in insects: structure, function and regulation of chitin synthases and chitinases},
  author={Merzendorfer, Hans and Zimoch, Lars},
  journal={Journal of Experimental Biology},
  volume={206},
  number={24},
  pages={4393--4412},
  year={2003},
  publisher={Company of Biologists}
}

@article{iwata2014real,
  title={Real-time in vivo imaging of butterfly wing development: revealing the cellular dynamics of the pupal wing tissue},
  author={Iwata, Masaki and Ohno, Yoshikazu and Otaki, Joji M},
  journal={PLoS One},
  volume={9},
  number={2},
  pages={e89500},
  year={2014},
  publisher={Public Library of Science San Francisco, USA}
}

@article{vermeulen1986chitin,
  title={Chitin biosynthesis by a fungal membrane preparation: Evidence for a transient non-crystalline state of chitin},
  author={Vermeulen, Cornelis A and Wessels, Joseph GH},
  journal={European Journal of Biochemistry},
  volume={158},
  number={2},
  pages={411--415},
  year={1986},
  publisher={Wiley Online Library}
}

@article{NORLEN2004,
title = {Stratum Corneum Keratin Structure, Function, and Formation: The Cubic Rod-Packing and Membrane Templating Model},
journal = {Journal of Investigative Dermatology},
volume = {123},
number = {4},
pages = {715-732},
year = {2004},
issn = {0022-202X},
doi = {https://doi.org/10.1111/j.0022-202X.2004.23213.x},
url = {https://www.sciencedirect.com/science/article/pii/S0022202X15309933},
author = {Lars Norlén and Ashraf Al-Amoudi}
}

@article{Ghiradella1974,
author = {Ghiradella, Helen},
title = {Development of ultraviolet-reflecting butterfly scales: How to make an interference filter},
journal = {Journal of Morphology},
volume = {142},
number = {4},
pages = {395-409},
doi = {https://doi.org/10.1002/jmor.1051420404},
url = {https://onlinelibrary.wiley.com/doi/abs/10.1002/jmor.1051420404},
eprint = {https://onlinelibrary.wiley.com/doi/pdf/10.1002/jmor.1051420404},
year = {1974}
}

@article{ParkerVignoliniCelluloseSelfAssAdvancedMater2018,
author = {Parker, Richard M. and Guidetti, Giulia and Williams, Cyan A. and Zhao, Tianheng and Narkevicius, Aurimas and Vignolini, Silvia and Frka-Petesic, Bruno},
title = {The Self-Assembly of Cellulose Nanocrystals: Hierarchical Design of Visual Appearance},
journal = {Advanced Materials},
volume = {30},
number = {19},
pages = {1704477},
doi = {https://doi.org/10.1002/adma.201704477},
url = {https://advanced.onlinelibrary.wiley.com/doi/abs/10.1002/adma.201704477},
eprint = {https://advanced.onlinelibrary.wiley.com/doi/pdf/10.1002/adma.201704477},
year = {2018}
}

@article{hou2021understanding,
  title={Understanding the structural diversity of chitins as a versatile biomaterial},
  author={Hou, Jiaxin and Aydemir, Berk Emre and Dumanli, Ahu G{\"u}mrah},
  journal={Philosophical Transactions of the Royal Society A},
  volume={379},
  number={2206},
  pages={20200331},
  year={2021},
  publisher={The Royal Society Publishing}
}

@article{binetti2009natural,
  title={The natural transparency and piezoelectric response of the \textit{Greta oto} butterfly wing},
  author={Binetti, Valerie R and Schiffman, Jessica D and Leaffer, Oren D and Spanier, Jonathan E and Schauer, Caroline L},
  journal={Integrative Biology},
  volume={1},
  number={4},
  pages={324--329},
  year={2009},
  publisher={Oxford University Press}
}

@article{schiffman2009solid,
  title={Solid state characterization of $\alpha$-chitin from \textit{Vanessa cardui Linnaeus} wings},
  author={Schiffman, Jessica D and Schauer, Caroline L},
  journal={Materials Science and Engineering: C},
  volume={29},
  number={4},
  pages={1370--1374},
  year={2009},
  publisher={Elsevier}
}

@article{dufresne2009self,
  title={Self-assembly of amorphous biophotonic nanostructures by phase separation},
  author={Dufresne, Eric R and Noh, Heeso and Saranathan, Vinodkumar and Mochrie, Simon GJ and Cao, Hui and Prum, Richard O},
  journal={Soft Matter},
  volume={5},
  number={9},
  pages={1792--1795},
  year={2009},
  publisher={Royal Society of Chemistry}
}

@article{bauernfeind2024not,
  title={Not only a matter of disorder in \uppercase{I-WP} minimal surface-based photonic networks: Diffusive structural color in \textit{Sternotomis amabilis} longhorn beetles},
  author={Bauernfeind, Viola and Saranathan, Vinodkumar and Djeghdi, Kenza and Longo, Elena and Flenner, Silja and Greving, Imke and Steiner, Ullrich and Wilts, Bodo D},
  journal={Materials Today Advances},
  volume={23},
  pages={100524},
  year={2024},
  publisher={Elsevier}
}

@article{wilts2018evolutionary,
  title={Evolutionary-optimized photonic network structure in white beetle wing scales},
  author={Wilts, Bodo D and Sheng, Xiaoyuan and Holler, Mirko and Diaz, Ana and Guizar-Sicairos, Manuel and Raabe, J{\"o}rg and Hoppe, Robert and Liu, Shu-Hao and Langford, Richard and Onelli, Olimpia D and others},
  journal={Advanced materials},
  volume={30},
  number={19},
  pages={1702057},
  year={2018},
  publisher={Wiley Online Library}
}

@article{wilts_pigmentary_2014,
	title = {Pigmentary and photonic coloration mechanisms reveal taxonomic relationships of the {Cattlehearts} ({Lepidoptera}: {Papilionidae}: {Parides})},
	volume = {14},
	issn = {1471-2148},
	shorttitle = {Pigmentary and photonic coloration mechanisms reveal taxonomic relationships of the {Cattlehearts} ({Lepidoptera}},
	url = {https://doi.org/10.1186/s12862-014-0160-9},
	doi = {10.1186/s12862-014-0160-9},
	number = {1},
	urldate = {2024-06-20},
	journal = {BMC Evolutionary Biology},
	author = {Wilts, Bodo D. and IJbema, Natasja and Stavenga, Doekele G.},
	month = jul,
	year = {2014},
	pages = {160},
}

@article{schroder2011chiral,
  title={The chiral structure of porous chitin within the wing-scales of \textit{Callophrys rubi}},
  author={Schr{\"o}der-Turk, Gerd E and Wickham, S and Averdunk, Holger and Brink, Frank and Gerald, JD Fitz and Poladian, L and Large, MCJ and Hyde, ST},
  journal={Journal of Structural Biology},
  volume={174},
  number={2},
  pages={290--295},
  year={2011},
  publisher={Elsevier},
doi={10.1016/j.jsb.2011.01.004}
}

@article{saba2014absence,
  title={Absence of circular polarisation in reflections of butterfly wing scales with chiral gyroid structure},
  author={Saba, Matthias and Wilts, Bodo D and Hielscher, Johannes and Schr{\"o}der-Turk, Gerd E},
  journal={Materials Today: Proceedings},
  volume={1},
  pages={193--208},
  year={2014},
  publisher={Elsevier},
doi={10.1016/j.matpr.2014.09.023}
}

@article{KirkensgaardPNASStripedGyroids2014,
author = {Jacob J. K. Kirkensgaard  and Myfanwy E. Evans  and Liliana de Campo  and Stephen T. Hyde },
title = {Hierarchical self-assembly of a striped gyroid formed by threaded chiral mesoscale networks},
journal = {Proceedings of the National Academy of Sciences},
volume = {111},
number = {4},
pages = {1271-1276},
year = {2014},
doi = {10.1073/pnas.1316348111},
URL = {https://www.pnas.org/doi/abs/10.1073/pnas.1316348111},
eprint = {https://www.pnas.org/doi/pdf/10.1073/pnas.1316348111}}

@article{Hielscher:17,
author = {Johannes Hielscher and Caroline Pouya and Peter Vukusic and Gerd E. Schr\"{o}der-Turk},
journal = {Opt. Express},
number = {5},
pages = {5001--5017},
publisher = {Optica Publishing Group},
title = {Harmonic distortions enhance circular dichroism of dielectric single gyroids},
volume = {25},
month = {03},
year = {2017},
url = {https://opg.optica.org/oe/abstract.cfm?URI=oe-25-5-5001},
doi = {10.1364/OE.25.005001},
}

@ARTICLE{EvansHydeSkin2011,
  title     = "From three-dimensional weavings to swollen corneocytes",
  author    = "Evans, Myfanwy E and Hyde, Stephen T",
  journal   = "J. R. Soc. Interface",
  publisher = "The Royal Society",
  volume    =  8,
  number    =  62,
  pages     = "1274--1280",
  month     =  sep,
  year      =  2011,
  language  = "en"
}

@article{EvansRothPRL2014Skin,
  title = {Shaping the Skin: The Interplay of Mesoscale Geometry and Corneocyte Swelling},
  author = {Evans, Myfanwy E. and Roth, Roland},
  journal = {Phys. Rev. Lett.},
  volume = {112},
  issue = {3},
  pages = {038102},
  numpages = {5},
  year = {2014},
  month = {01},
  publisher = {American Physical Society},
  doi = {10.1103/PhysRevLett.112.038102},
  url = {https://link.aps.org/doi/10.1103/PhysRevLett.112.038102}
}

@article{wetzel2024triplyperiodichelicalweaves,
author = {Duston Wetzel and Paul Gailiunas and Moses Gaither-Ganim and William Holt},
journal = {arXiv},
pages = {2402.07849},
title = {Triply Periodic Helical Weaves},
year = {2024},
url = {https://arxiv.org/abs/2402.07849},
doi = {https://doi.org/10.48550/arXiv.2402.07849},
}

@article{PouyaOrderedAndDisorderedDiamond2011,
author = {C. Pouya and D. G. Stavenga and P. Vukusic},
journal = {Opt. Express},
number = {12},
pages = {11355--11364},
publisher = {Optica Publishing Group},
title = {Discovery of ordered and quasi-ordered photonic crystal structures in the scales of the beetle \textit{Eupholus magnificus}},
volume = {19},
month = {06},
year = {2011},
url = {https://opg.optica.org/oe/abstract.cfm?URI=oe-19-12-11355},
doi = {10.1364/OE.19.011355},
}

@ARTICLE{McNamaraDiamondFossil2014,
  title     = "Cryptic iridescence in a fossil weevil generated by single
               diamond photonic crystals",
  author    = "McNamara, Maria E and Saranathan, Vinod and Locatelli, Emma R
               and Noh, Heeso and Briggs, Derek E G and Orr, Patrick J and Cao,
               Hui",
  journal   = "J. R. Soc. Interface",
  publisher = "The Royal Society",
  volume    =  11,
  number    =  100,
  pages     = "20140736",
  month     =  nov,
  year      =  2014,
  language  = "en"
}

@article{WiltsDiamondBeetle2012,
author = {Wilts, Bodo D.  and Michielsen, Kristel  and Kuipers, Jeroen  and De Raedt, Hans  and Stavenga, Doekele G. },
title = {Brilliant camouflage: photonic crystals in the diamond weevil, \textit{Entimus imperialis}},
journal = {Proceedings of the Royal Society B: Biological Sciences},
volume = {279},
number = {1738},
pages = {2524-2530},
year = {2012},
doi = {10.1098/rspb.2011.2651},

URL = {https://royalsocietypublishing.org/doi/abs/10.1098/rspb.2011.2651},
eprint = {https://royalsocietypublishing.org/doi/pdf/10.1098/rspb.2011.2651}
}

@article{Saranathan2021,
  title = {Evolution of single gyroid photonic crystals in bird feathers},
  volume = {118},
  ISSN = {1091-6490},
  url = {http://dx.doi.org/10.1073/pnas.2101357118},
  DOI = {10.1073/pnas.2101357118},
  number = {23},
  journal = {Proceedings of the National Academy of Sciences},
  publisher = {Proceedings of the National Academy of Sciences},
  author = {Saranathan,  Vinodkumar and Narayanan,  Suresh and Sandy,  Alec and Dufresne,  Eric R. and Prum,  Richard O.},
  year = {2021},
  month = may 
}

@article{michielsen2008gyroid,
  title={Gyroid cuticular structures in butterfly wing scales: biological photonic crystals},
  author={Michielsen, Kristel and Stavenga, Doekele G},
  journal={Journal of The Royal Society Interface},
  volume={5},
  number={18},
  pages={85--94},
  year={2008},
  publisher={The Royal Society London},
doi={10.1098/rsif.2007.1065}
}

@article{SabaPRL2011,
  title = {Circular dichroism in biological photonic crystals and cubic chiral nets},
  author = {Saba, M. and Thiel, M. and Turner, M. D. and Hyde, S. T. and Gu, M. and Grosse-Brauckmann, K. and Neshev, D. N. and Mecke, K. and Schr\"oder-Turk, G. E.},
  journal = {Phys. Rev. Lett.},
  volume = {106},
  issue = {10},
  pages = {103902},
  numpages = {4},
  year = {2011},
  month = {03},
  publisher = {American Physical Society},
  doi = {10.1103/PhysRevLett.106.103902},
  url = {https://link.aps.org/doi/10.1103/PhysRevLett.106.103902}
}

@article{wilts2012iridescence,
  title={Iridescence and spectral filtering of the gyroid-type photonic crystals in \textit{Parides sesostris} wing scales},
  author={Wilts, Bodo D. and Michielsen, Kristel. and De Raedt, Hans. and Stavenga, Doekele G},
  journal={Interface Focus},
  volume={2},
  number={5},
  pages={681--687},
  year={2012},
  publisher={The Royal Society},
doi={10.1098/rsfs.2011.0082}
}

@article{day2019sub,
  title={Sub-micrometer insights into the cytoskeletal dynamics and ultrastructural diversity of butterfly wing scales},
  author={Day, Christopher R and Hanly, Joseph J and Ren, Anna and Martin, Arnaud},
  journal={Developmental Dynamics},
  volume={248},
  number={8},
  pages={657--670},
  year={2019},
  publisher={Wiley Online Library},
doi={10.1002/dvdy.63}
}

@article{seah2023hierarchical,
  title={Hierarchical morphogenesis of swallowtail butterfly wing scale nanostructures},
  author={Seah, Kwi Shan and Saranathan, Vinodkumar},
  journal={Elife},
  volume={12},
  pages={RP89082},
  year={2023},
  publisher={eLife Sciences Publications Limited},
doi= {10.7554/eLife.89082.3}

}

@article{wilts2019nature,
  title={Nature’s functional nanomaterials: Growth or self-assembly?},
  author={Wilts, Bodo D and Clode, Peta L and Patel, Nipam H and Schr{\"o}der-Turk, Gerd E},
  journal={MRS Bulletin},
  volume={44},
  number={2},
  pages={106--112},
  year={2019},
  publisher={Cambridge University Press},
doi={10.1557/mrs.2019.21}
}

@article{ghiradella1989structure,
  title={Structure and development of iridescent butterfly scales: lattices and laminae},
  author={Ghiradella, H},
  journal={Journal of Morphology},
  volume={202},
  number={1},
  pages={69--88},
  year={1989},
  publisher={Wiley Online Library},
doi={10.1002/jmor.1052020106}
}

@article{lloyd2024actin,
  title={The actin cytoskeleton plays multiple roles in structural colour formation in butterfly wing scales},
  author={Lloyd, Victoria J and Burg, Stephanie L and Harizanova, Jana and Garcia, Esther and Hill, Olivia and Enciso-Romero, Juan and Cooper, Rory L and Flenner, Silja and Longo, Elena and Greving, Imke and others},
  journal={Nature Communications},
  volume={15},
  number={1},
  pages={4073},
  year={2024},
  publisher={Nature Publishing Group UK London},
doi={10.1038/s41467-024-48060-3}
}

@article{McDougalKolle2021PNAS,
author = {Anthony D. McDougal  and Sungsam Kang  and Zahid Yaqoob  and Peter T. C. So  and Mathias Kolle },
title = {\textit{In vivo} visualization of butterfly scale cell morphogenesis in \textit{Vanessa cardui}},
journal = {Proceedings of the National Academy of Sciences},
volume = {118},
number = {49},
pages = {e2112009118},
year = {2021},
doi = {10.1073/pnas.2112009118},
URL = {https://www.pnas.org/doi/abs/10.1073/pnas.2112009118},
eprint = {https://www.pnas.org/doi/pdf/10.1073/pnas.2112009118}}

@article{WiltsScienceAdv2017,
author = {Bodo D. Wilts  and Benjamin Apeleo Zubiri  and Michael A. Klatt  and Benjamin Butz  and Michael G. Fischer  and Stephen T. Kelly  and Erdmann Spiecker  and Ullrich Steiner  and Gerd E. Schröder-Turk },
title = {Butterfly gyroid nanostructures as a time-frozen glimpse of intracellular membrane development},
journal = {Science Advances},
volume = {3},
number = {4},
pages = {e1603119},
year = {2017},
doi = {10.1126/sciadv.1603119},
URL = {https://www.science.org/doi/abs/10.1126/sciadv.1603119},
eprint = {https://www.science.org/doi/pdf/10.1126/sciadv.1603119}}

@article{Saranathan2010,
author = {Vinodkumar Saranathan  and Chinedum O. Osuji  and Simon G. J. Mochrie  and Heeso Noh  and Suresh Narayanan  and Alec Sandy  and Eric R. Dufresne  and Richard O. Prum },
title = {Structure, function, and self-assembly of single network gyroid (\textit{I}$4_1$32) photonic crystals in butterfly wing scales},
journal = {Proceedings of the National Academy of Sciences},
volume = {107},
number = {26},
pages = {11676-11681},
year = {2010},
doi = {10.1073/pnas.0909616107},
URL = {https://www.pnas.org/doi/abs/10.1073/pnas.0909616107},
eprint = {https://www.pnas.org/doi/pdf/10.1073/pnas.0909616107}}

@ARTICLE{Dinwiddie2014,
  title     = "Dynamics of F-actin prefigure the structure of butterfly wing
               scales",
  author    = "Dinwiddie, April and Null, Ryan and Pizzano, Maria and Chuong,
               Lisa and Leigh Krup, Alexis and Ee Tan, Hwei and Patel, Nipam H",
  journal   = "Developmental Biology",
  publisher = "Elsevier BV",
  volume    =  392,
  number    =  2,
  pages     = "404--418",
  month     =  aug,
  year      =  2014,
  copyright = "https://www.elsevier.com/open-access/userlicense/1.0/",
doi={10.1016/j.ydbio.2014.06.005}

}

\newpage

\setcounter{figure}{0}
\renewcommand{\figurename}{Figure}
\renewcommand{\thefigure}{S\arabic{figure}}

\section*{Supporting Information}

\begin{figure}[h!]
\centering
\includegraphics[width=\columnwidth]{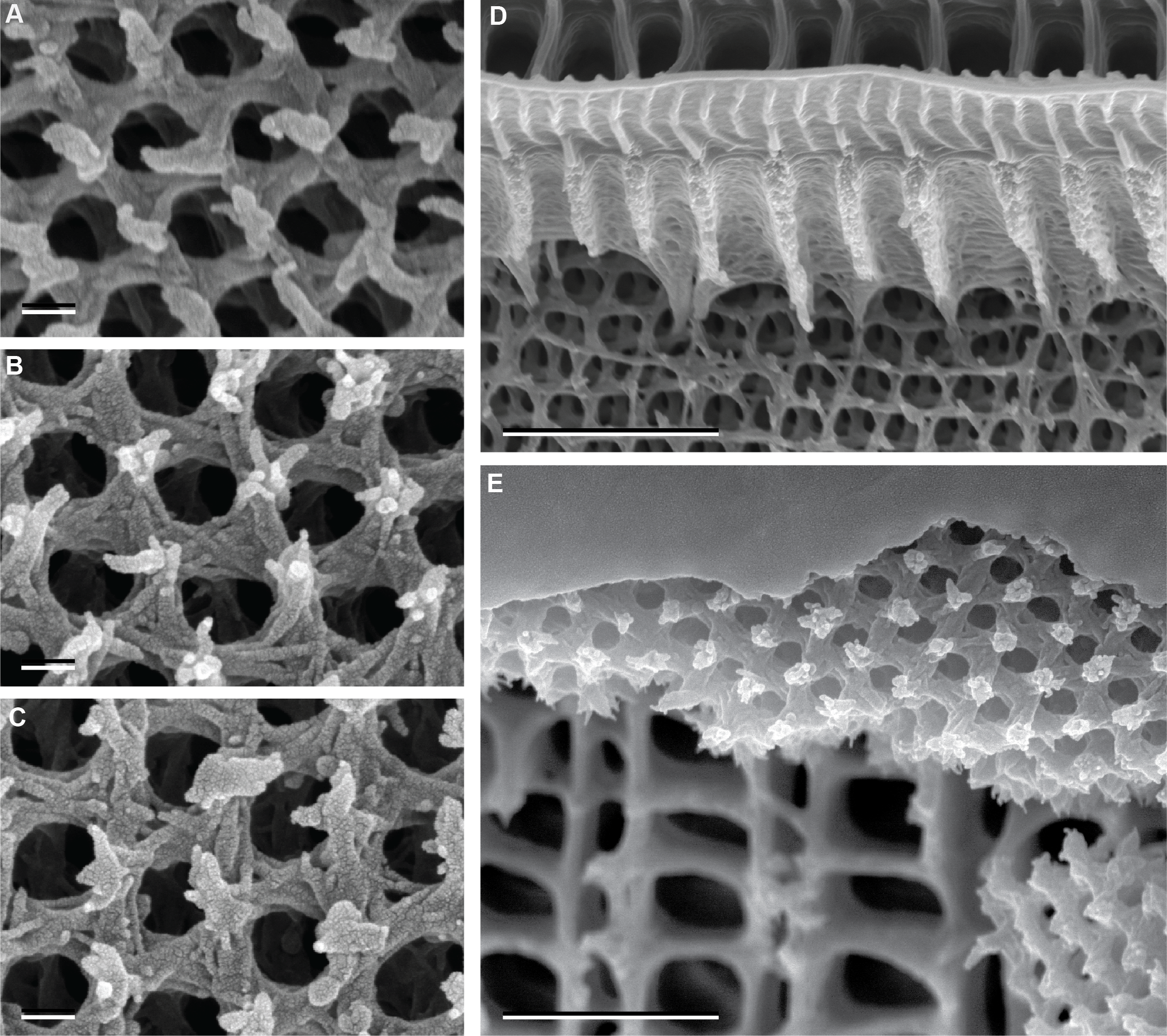}
\caption{SEM images of critical-point dried samples demonstrating the variation in fibre numbers. (A--C) SEM images of samples taken at day 12 (A), day 13 (B), and day 15 (C), showing a change in the number of entangled fibres throughout development. (D, E) The fibrous gyroid network adjacent to the upper lamina (D) is comprised of fewer fibres than the network adjacent to the lower lamina (E). }\label{fig:sem_images}
\end{figure}

\FloatBarrier

\begin{figure}[t]
\centering
\includegraphics[width=\columnwidth]{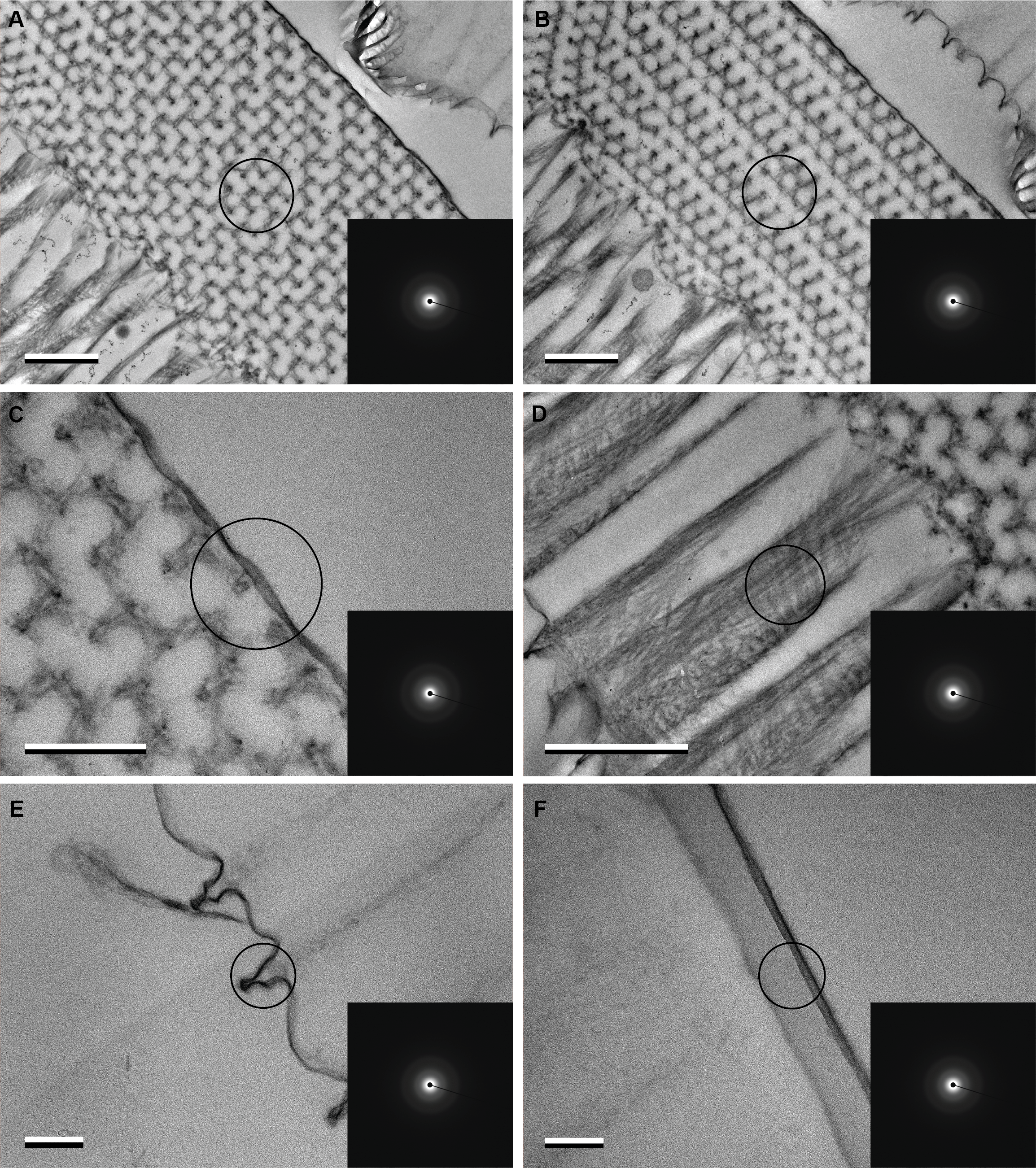}
\caption{TEM and SAED data of \textit{Parides sesostris} samples fixed at day 15: (A,B) Green gyroid-containing scale cells with the diffraction mask set to regions that only contain gyroid fibres. (C,D) Green gyroid-containing scale cells with the diffraction mask set to include the lower lamina and the diffuser, respectively. (E,F) Black ground scale cells, with the mask focused on the upper ridges and the lower lamina. Analyses were carried out on resin embedded, 2\% osmium tetroxide stained samples that were sectioned to a thickness of 250 nm. The black circle in the TEM images indicates the aperture (mask) used for the diffraction analysis. The insets show the scattering pattern obtained for each measurement. Scale bars = 1$\mu$m (A,B,D), 500 nm (C), and 200 nm (E,F).}\label{fig:electron_diffraction}
\end{figure}

\begin{figure}[t]
\centering
\includegraphics[width=\textwidth]{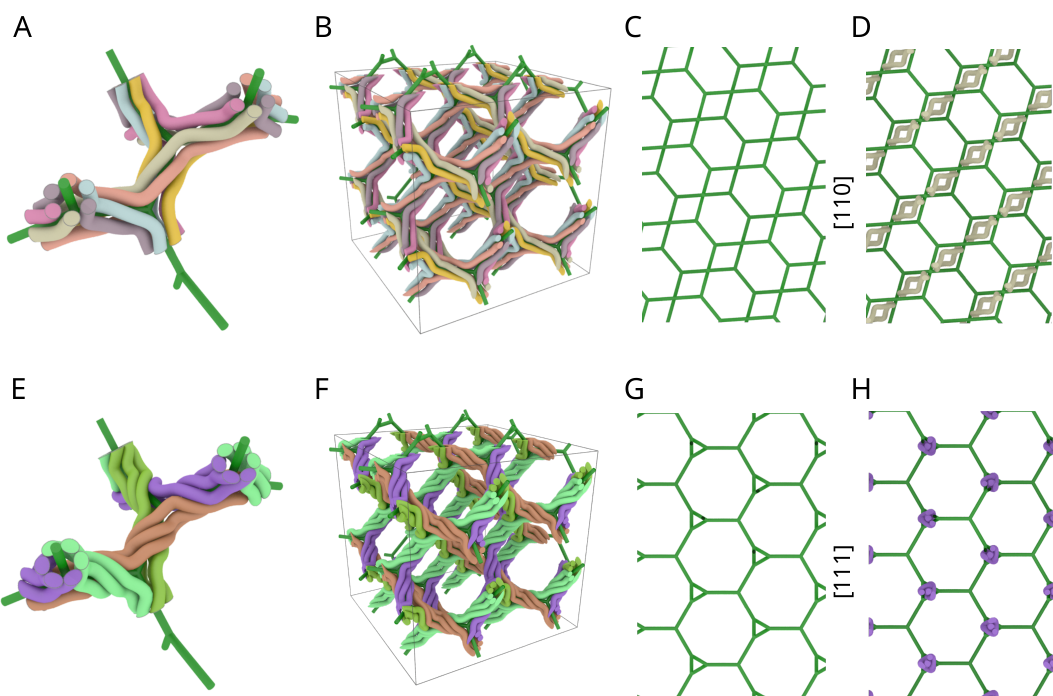}\\

\caption{Comparison of two geometric models $(\frac{0.8}{6})^6$ (A--D) and $(\frac{1.8}{6})^6$ (E--H) for 6-threaded weavings around the \texttt{srs} graph of the gyroid: (A,B) Small section and $2^3$ translational unit cells of the $(\frac{0.8}{6})^6$ model; (C) the \texttt{srs} graph projected onto the [110] plane; (D) The $(\frac{0.8}{6})^6$ model breaks up into individual deformed helices along the six [110] directions; this image shows all those helices (in cream) that run along the [110] direction perpendicular to the paper. (E,F) Small section and $2^3$ translational unit cells of the $(\frac{1.8}{6})^6$ model; (G) the \texttt{srs} graph projected onto the [111] plane; (H) The $(\frac{1.8}{6})^6$ model breaks up into individual triplets of deformed helices along the 4 [111] directions; this image shows all those helices (in purple) that run along the [111] direction perpendicular to the paper; each purple point corresponds to a bundle of three (triplet) helices revolving around the corresponding [111] axis. }\label{suppfig:geometric-model-compare-twist-0-and-1}
\end{figure}

\begin{figure}[t]
\centering
\includegraphics[width=12cm]{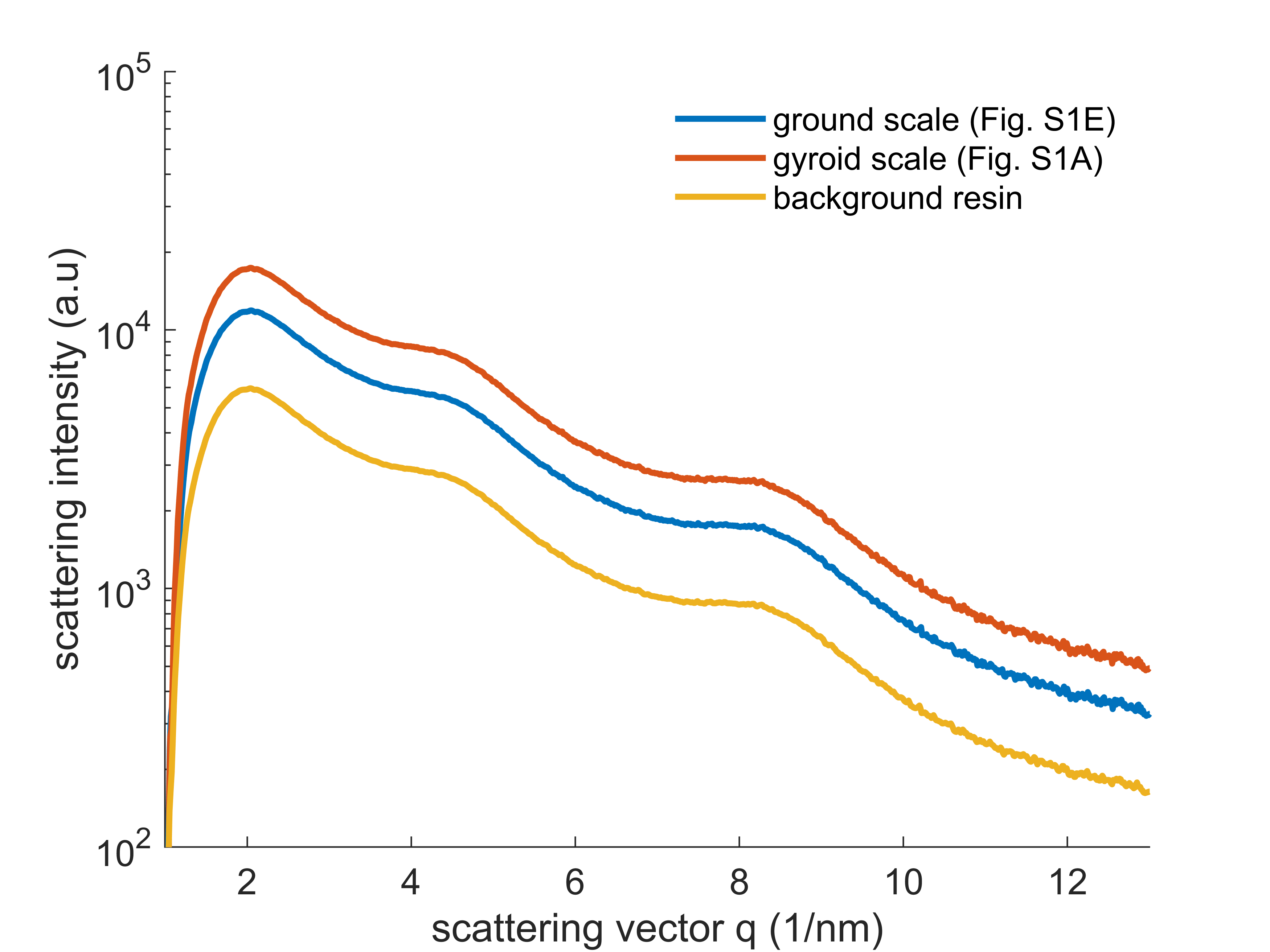}
\caption{Azimuthally averaged scattering intensities of the electron diffraction data of the measurements shown in Supplementary Figure \ref{fig:electron_diffraction}, of fixed tissue of the developing butterfly at day 15. 
Measurements are shown for the data of a ground scale (Fig.\ S1E), of a gyroid forming scale (Fig.\ S1A) and of the background, that is, of a sample that only contains the resin used in these experiments. This presentation of the data provides further demonstration that the butterfly diffraction data contain no discernible features beyond the resin background. 
}
\label{fig:line_profile_ed}
\end{figure}

\begin{figure}[t]
\centering
\includegraphics[width=0.48\linewidth]{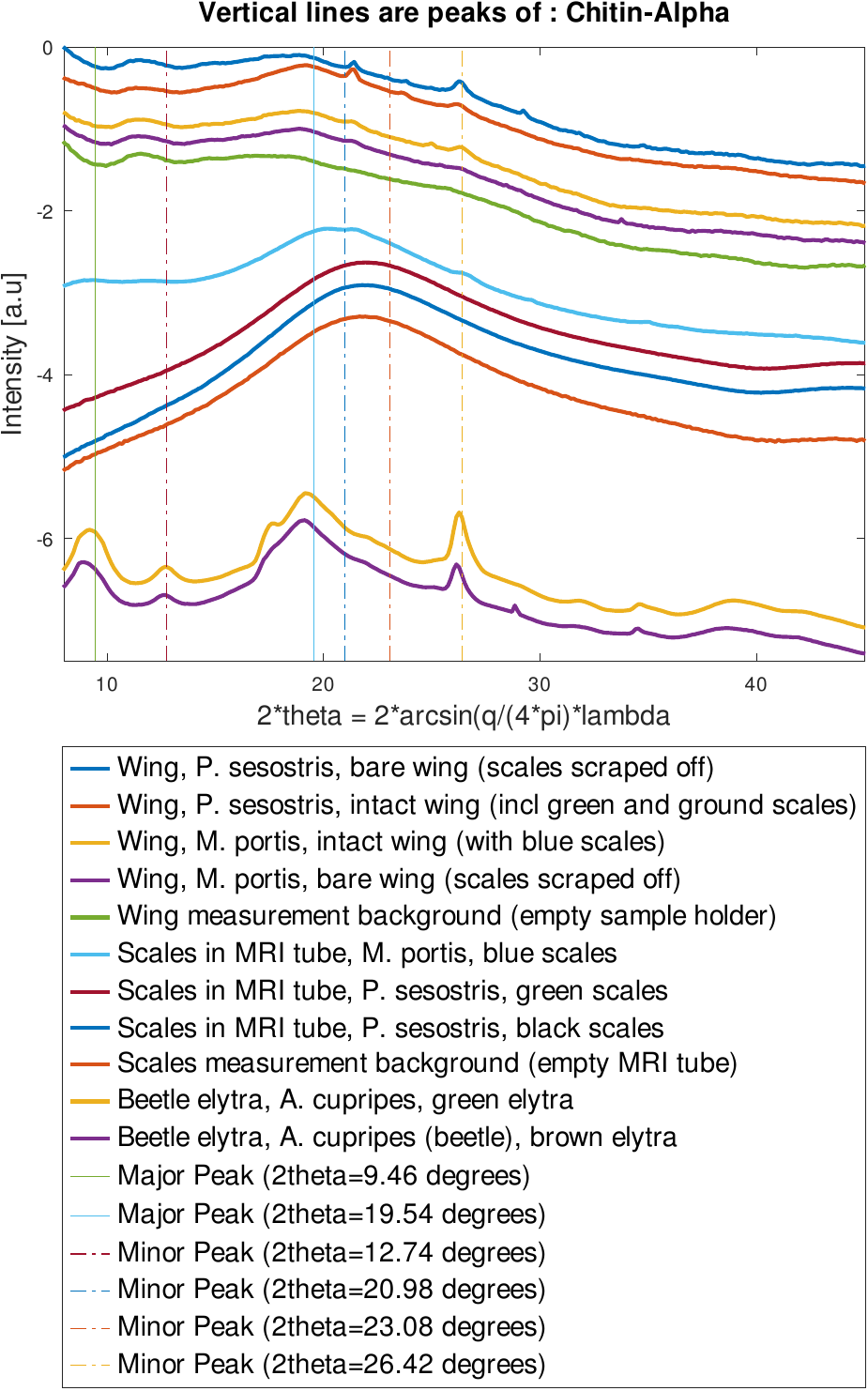}\nolinebreak\hfill
\includegraphics[width=0.48\linewidth]{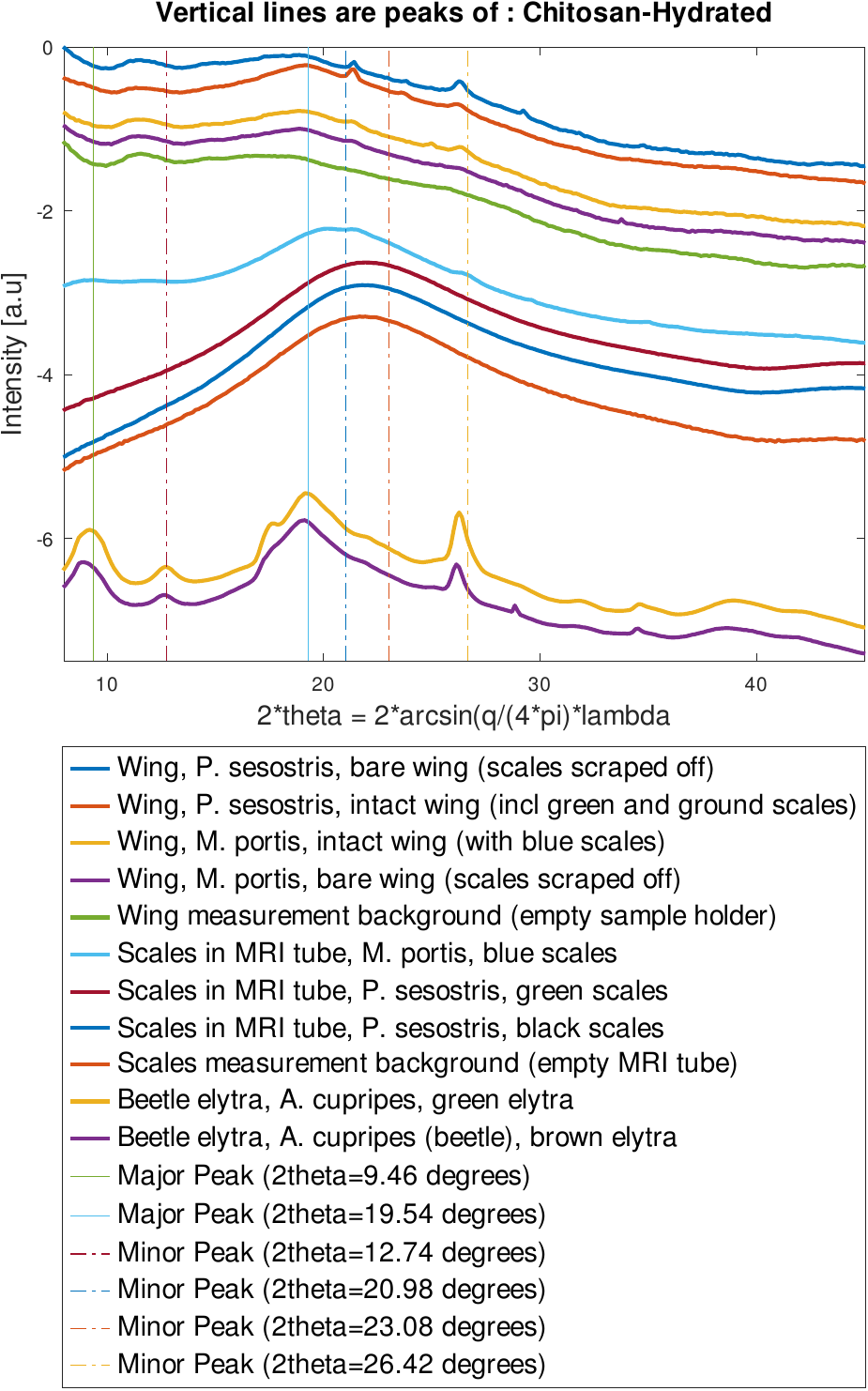}\\
\vspace*{0.5cm}

\caption{Azimuthally averaged X-ray scattering intensity of sections of butterfly wings (with and without scales removed), of a powder sample of intact butterfly wing scales in an MRI tube and, as a reference, of sections of the elytra of \textit{Anomala cupripes} (green cuticle and brown cuticle). All were obtained from dried specimens of the adult animals that had undergone normal development. The \textit{Parides sesostris} scales were collected from circa 30 specimens and were a mixture of butterflies collected in Panama as part of this project, and specimens purchased from World of Butterflies (wobam.co.uk) labelled as originating from Peru, and of \textit{Morpho portis} specimens. Specimens of \textit{M.\ portis} were samples previously purchased from an unknown commercial insect supplier. The beetle specimens of \textit{A.\ cupripes} were added as a reference for a biological material with a Bouligand reflector structure. Intensities are in arbitrary units on a logarithmic scale, with curves shifted up or down to enable better visibility. Data was compared to all chitin and chitosan phases described in Table 1 of \cite{Tsurkan2021}. The two plots are the same scattering data, and differ only in the vertical line that indicate the major and minor peaks of 'alpha chitin' (left) and 'chitosan hydrated' (right). Among the phases discussed in \cite{Tsurkan2021} only these two  provide reasonably similar peak positions to the peaks of the beetle data. The powder samples of scales of \textit{P.\ sesostris} show no indication of any crystalline order. The powder sample of scales of \textit{M.\ portis} shows some tiny peaks that may relate to an underlying small crystalline contribution of the constituent material, but are not clearly related to a chitin phase. The wing samples for both butterfly species, both with and without scales, show some peaks which however do not clearly relate to a chitin phase. (For identification in black/white print: legends are in order of the vertical position of the scattering curves, with the bare wing sample of \textit{P.\ sesostris} the top curve and the brown elytra of \textit{A.\ cupripes} at the bottom.)  }\label{fig:saxs-waxs-copenhagen-butterfly-good-fit-phase}
\end{figure}

\begin{figure}[t]
\centering
\includegraphics[width=\linewidth]{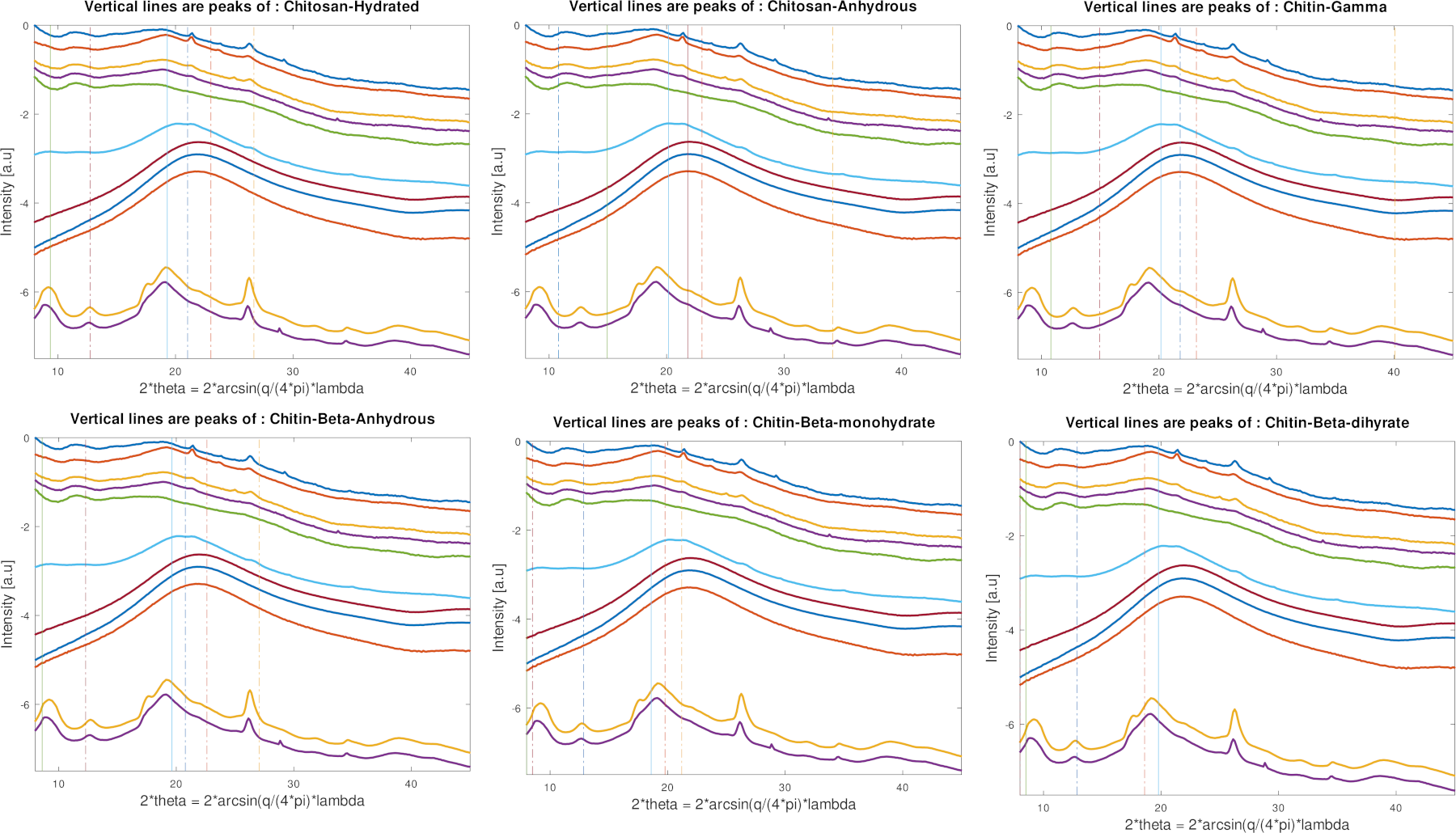}\\
\vspace*{0.5cm}
\caption{The same X-ray scattering data as in Fig.\ \ref{fig:saxs-waxs-copenhagen-butterfly-good-fit-phase}, presented together with vertical lines representing other, less well fitting, phases as per \cite{Tsurkan2021}. See Fig.\ \ref{fig:saxs-waxs-copenhagen-butterfly-good-fit-phase} for legends.}\label{fig:saxs-waxs-copenhagen-butterfly-poor-fit-phase}
\end{figure}

\FloatBarrier
\pagebreak
\clearpage

\begin{figure}[t]
\centering
\includegraphics[width=14cm]{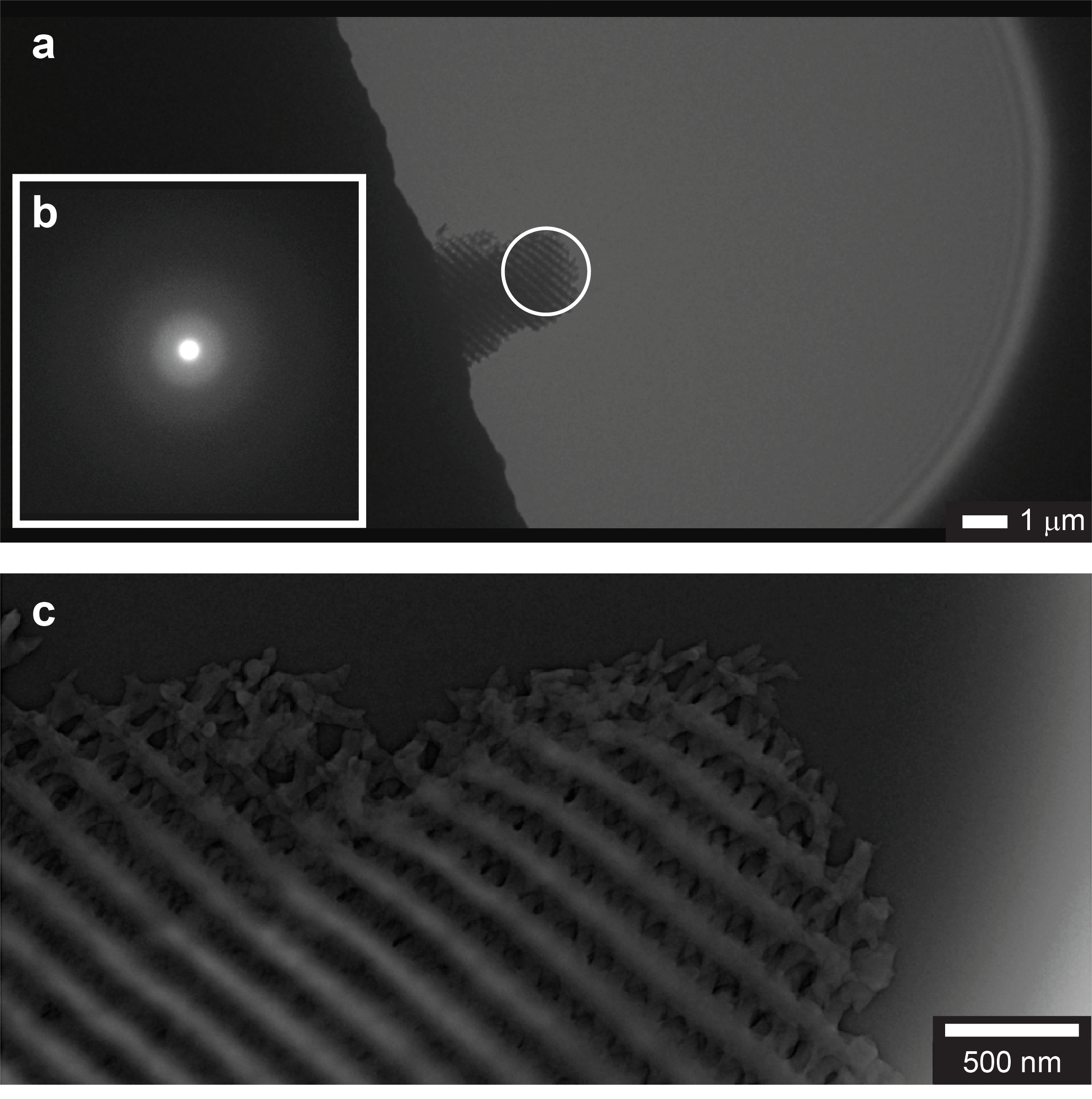}

\caption{Electron diffraction analysis carried out on gyroid-containing fragments in mature dry \textit{P.\ sesostris} wing scales: a typical example of a gyroid fragment found in the dedicated electron diffractometer, after green scales from mature butterflies were gently ground between two glass slides. Two real-space images are presented (a and c) of the same fragment (at different magnifications), and the inset (b) shows the image acquired when the optics were switched to project diffracted waves onto the detector (showing only the primary beam and diffuse scattering). The white circle in panel (a) represents the area selected during the diffraction imaging (b). }\label{fig:ed-dtu-mature-scale-gyroid-frag-no-diffraction}
\end{figure}

\begin{figure}[t]
\centering
\includegraphics[width=12cm]{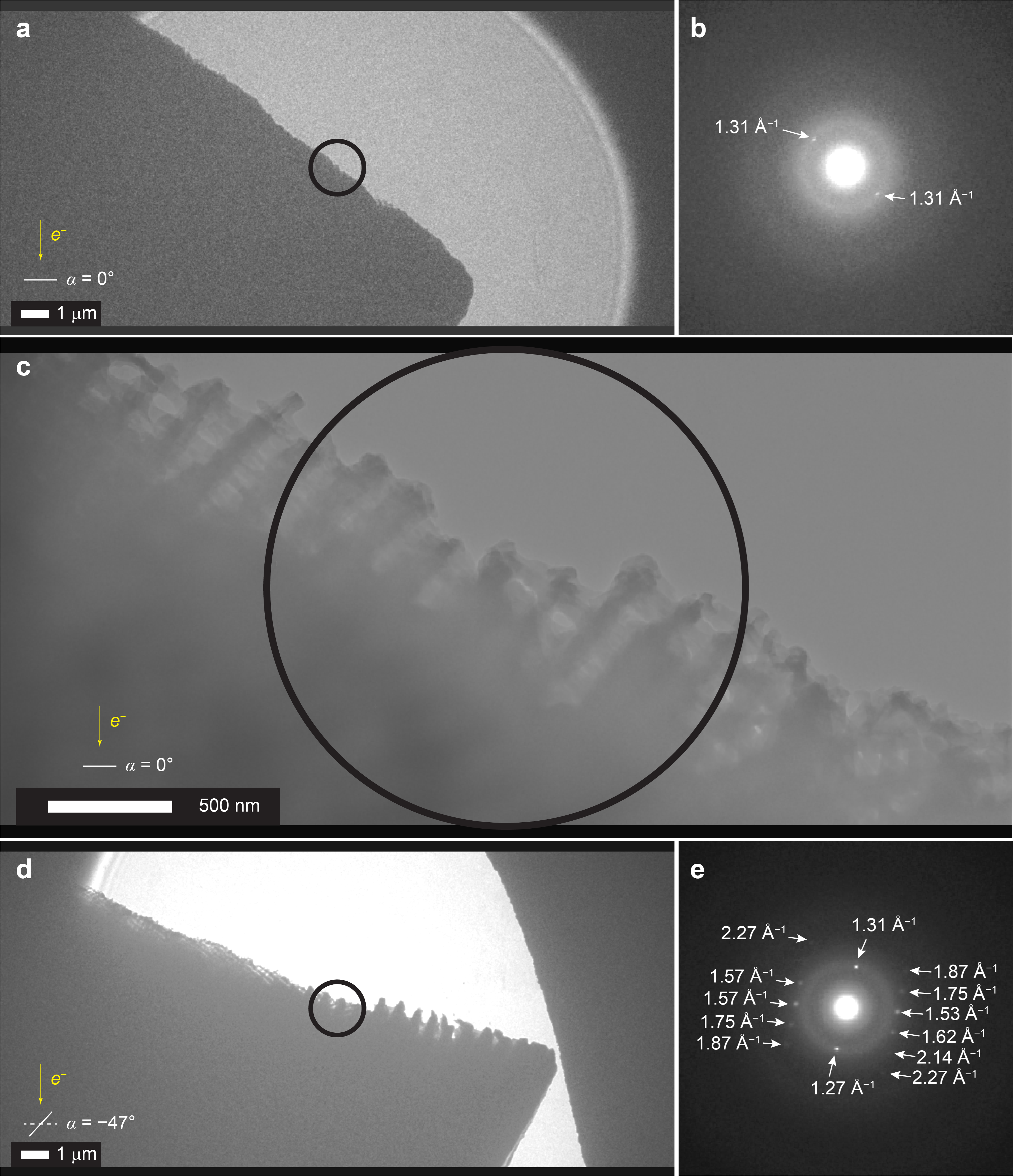}\\ \vspace*{0.3cm}
\includegraphics[width=12cm]{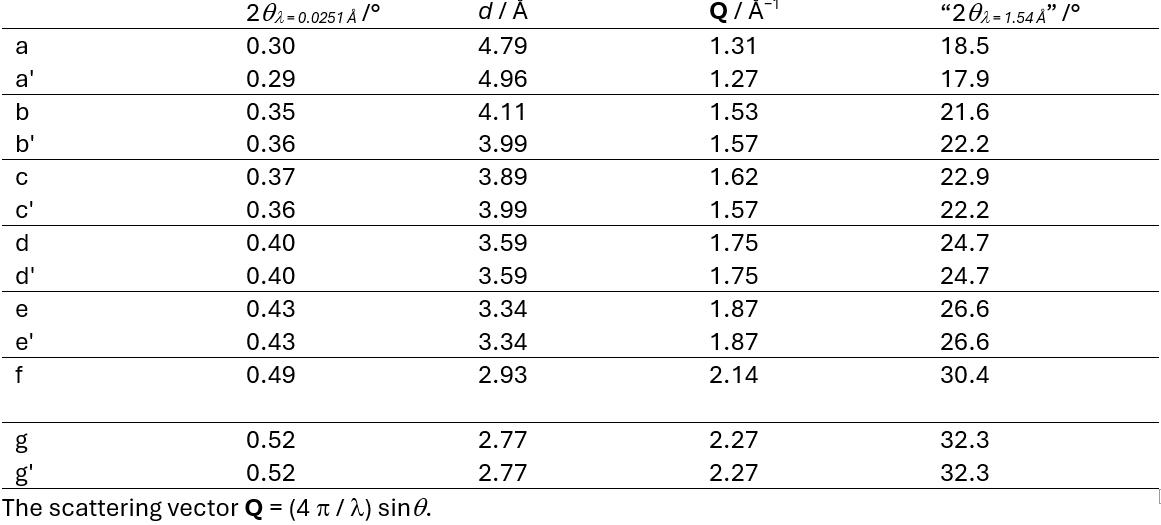}

\caption{Electron diffraction analysis carried out on gyroid-containing intact (but broken) scales in mature dry \textit{P.\ sesostris} wing scales: images of one of the very few instances of gyroid-containing green scales from a mature \textit{P.\ sesostris} wing which showed fleeting diffraction at $\alpha \approx 47^o$. Panels a and c show the scale at neutral tilt ($\alpha\approx 0^o$), with the diffraction image acquired at this angle shown in panel (b). Panel (d) shows a real-space image of the scale at  $\alpha \approx 47^o$, where a number of Bragg peaks were observed in the diffraction image (e). In the real-space images (panels a, c and d), the area selected during the diffraction imaging is shown by black circles, and in the diffraction images (panels b and e) the scattering vectors, \textbf{Q} for weak Bragg peaks are annotated. The table below shows the $2\theta$ angles for the reflections observed in panel (e), the equivalent $d$ spacing (\AA) and scattering vectors (Q / $\AA^{-1}$), and the corresponding $2\theta$ angles for $\lambda = 1.54 \AA$ (as used in the X-ray Scattering studies).}\label{fig:ed-dtu-mature-scale-gyroid-intactscale-some-scatter}
\end{figure}

\FloatBarrier

\begin{table}[h!]
\includegraphics[width=\textwidth]{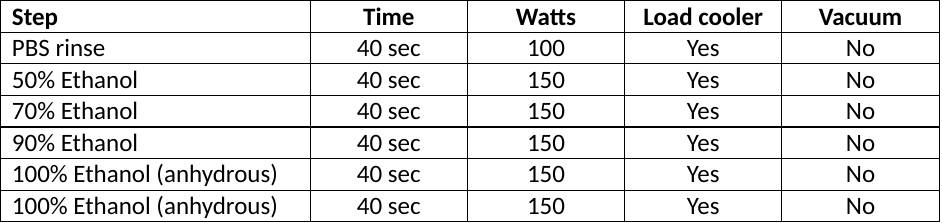} \caption{PELCO Biowave protocol for SEM sample preparation.} \label{supptable:sem-pelco}
\end{table}

\FloatBarrier

\begin{table}
\includegraphics[width=\textwidth]{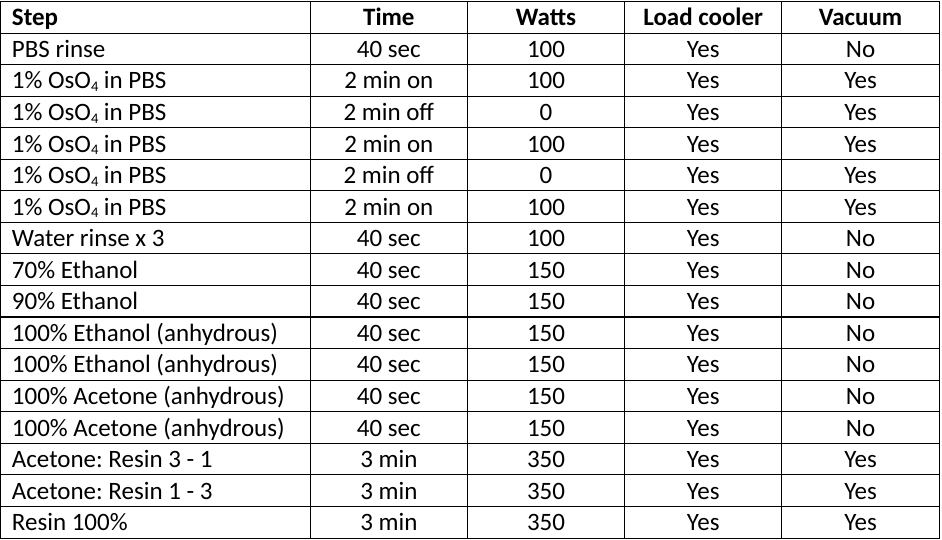} \caption{PELCO Biowave protocol for TEM sample preparation.} \label{supptable:tem-pelco} 
\end{table}

\end{document}